\documentclass[aps,prl,reprint,preprintnumbers]{revtex4-2}
\usepackage{blindtext}
\usepackage{chngcntr}
\counterwithout{equation}{section} 
\usepackage{graphicx}
\usepackage{bbm}
\usepackage{amsmath}
\usepackage{amssymb}
\usepackage{mathtools}
\usepackage{amsthm}
\usepackage{hyperref}
\usepackage{mathrsfs}
\usepackage{marvosym}
\usepackage{dsfont}
\usepackage{xcolor}
\usepackage{braket}
\usepackage{cleveref}
\usepackage{comment}
\usepackage{float}
\usepackage{fancybox}
\usepackage[skins,theorems]{tcolorbox}
\tcbset{highlight math style={enhanced,
		colframe=black,colback=white,arc=0pt,boxrule=1pt}}

\newcommand{\beq}{\begin{equation}}  \newcommand{\eeq}{\end{equation}}
\newcommand{\bal}{\begin{aligned}}   \newcommand{\eal}{\end{aligned}}
\def\beqa{\begin{eqnarray}}
	\def\eeqa{\end{eqnarray}}

\newcommand{\cF}{\mathcal{F}}

\newcommand{\Mpl}{M_{\textrm{Pl}}}
\def\Mpl{M_{\text{Pl}}}

\def\CL {{\cal L}}
\def\CN {{\cal N}}

\def\CK {{\cal K}}
\def\CV {{\cal V}}

\def\CF {{\cal F}}
\def\CR {{\cal R}}
\def\CQ {{\cal Q}}
\def\CI {{\cal I}}
\def\CD {{\cal D}}

\def\p {{\partial}}
\def\be{\begin{equation}}
\def\ee{\end{equation}}
\def\bea{\begin{eqnarray}}
\def\eea{\end{eqnarray}}
\def\bes{\begin{subequations}}
\def\ees{\end{subequations}}

\def\oh{\frac{1}{2}}

\def\re{\mbox{Re}\, }
\def\im{\mbox{Im}\, }

\def\id{\mathbbm{1}}

\def\simleq{\; \raise0.3ex\hbox{$<$\kern-0.75em
		\raise-1.1ex\hbox{$\sim$}}\; }
\def\simgeq{\; \raise0.3ex\hbox{$>$\kern-0.75em
		\raise-1.1ex\hbox{$\sim$}}\; }

\theoremstyle{remark}

\newtheoremstyle{named}{}{}{\itshape}{}{\bfseries}{.}{.5em}{#3}
\theoremstyle{named}

\hypersetup{
	colorlinks = true,
	citecolor = {red},
	linkcolor = {blue},
	urlcolor = {blue},
}

\begin{document}
\title{The Moduli Space Curvature and the Weak Gravity Conjecture}
\author{Alberto Castellano}
\email{acastellano@uchicago.edu}
    \affiliation{Enrico Fermi Institute \& Kadanoff Center for Theoretical Physics, University of Chicago, Chicago, IL 60637, USA}
    \affiliation{Kavli Institute for Cosmological Physics, University of Chicago, Chicago, IL 60637, USA}
\author{Fernando Marchesano} 
\email{fernando.marchesano@csic.es}
\author{Luca Melotti} 
\email{luca.melotti@ift.csic.es}
\author{Lorenzo Paoloni} 
\email{lorenzo.paoloni@csic.es}
\affiliation{Instituto de F\'{\i}sica Te\'orica UAM-CSIC, c/ Nicol\'as Cabrera 13-15, 28049 Madrid, Spain}
\affiliation{Departamento de F\'{\i}sica Te\'orica, Universidad Aut\'onoma de Madrid, 28049 Madrid, Spain}
	
\begin{abstract}
We unveil a remarkable interplay between rigid field theories (RFTs), charge-to-mass ratios $\gamma$ and scalar curvature divergences $\mathsf{R}_{\rm div}$ in the vector multiplet moduli space of 4d ${\cal N}=2$ supergravities, obtained upon compactifying type II string theory on Calabi--Yau threefolds. We show that the condition to obtain an RFT that decouples from gravity implies a divergence in the $\gamma$ of (would-be) BPS particles charged under the rigid theory, and vice-versa. For weak coupling limits, where the scalar curvature diverges, we argue that such BPS particles exist and that $\mathsf{R}_{\rm div} \lesssim \gamma^2$, implying that all these divergences are a consequence of RFT limits. More precisely, along geodesics we find that $\mathsf{R}_{\rm div} \sim (\Lambda_{\rm wgc}/\Lambda_g)^2$, where $\Lambda_{\rm wgc} \equiv g_{\rm rigid} \Mpl$ is the RFT cut-off estimate of the Weak Gravity Conjecture and $\Lambda_g = g_{\rm rigid}^{-2} \Lambda_{\rm RFT}$ the electrostatic energy integrated up to its actual cut-off $\Lambda_{\rm RFT}$.

\end{abstract}
\preprint{EFI-24-10, IFT-UAM/CSIC-24-145}
\maketitle

\section{Introduction}
A guiding principle behind the Swampland Programme \cite{Vafa:2005ui} is the idea that for Effective Field Theories (EFTs) consistent with Quantum Gravity (QG) there are non-trivial relations between UV and IR data. Examples of this are the magnetic version of the Weak Gravity Conjecture (WGC) \cite{Arkani-Hamed:2006emk}, which predicts the maximal cut-off $\Lambda_{\rm wgc} = g \Mpl$ for a $U(1)$ gauge interaction with coupling $g$,  and the characterisation of the species scale $\Lambda_{\rm sp}$ in terms of IR quantities, see e.g. \cite{Castellano:2024bna} and references therein. 

Here we focus our attention on a piece of data that is oftentimes neglected, namely the EFT field space curvature. More precisely, we consider the curvature of four-dimensional $\CN=2$ vector multiplet moduli spaces obtained from type II string theory compactified on Calabi--Yau threefolds. This is a particularly interesting framework in string theory. On the one hand, it provides a rich set of 4d EFTs compatible with Quantum Gravity, which display non-trivial relations between gravity and gauge interactions. On the other hand, their field space metric is known exactly and, in many instances, it is simple enough for questions to be addressed analytically. 

It is precisely in this setup that the field space scalar curvature has been lately analysed in \cite{Marchesano:2023thx,Marchesano:2024tod,CMP}. The emerging picture so far is that divergences in the scalar curvature $\mathsf{R}$ signal the decoupling of a rigid field theory (RFT) from gravity, both along weak-coupling trajectories of finite and infinite distance, or duals thereof. Moreover, for type IIA large-volume limits the degree of divergence in terms of the quotient $\Lambda_{\rm wgc}/\Lambda_{\rm sp}$ -- where $\Lambda_{\rm wgc} \equiv g_{\rm rigid} \Mpl$ with $g_{\rm rigid}$  the RFT gauge coupling -- indicates to which kind of higher-dimensional strongly-coupled rigid theory the RFT flows unto in the UV  \cite{Marchesano:2024tod}. This already shows that the IR data involved in the moduli space scalar curvature also contains non-trivial UV information.

In this letter we would like to revisit these recent results from a single, overarching perspective, that not only unifies the way in which curvature divergences and RFT limits are related, but also gives further insight on the UV/IR interplay associated to them. The key object in this story will be the charge-to-mass ratio $\gamma$ of BPS particles, made up from D-branes wrapping internal cycles of the Calabi--Yau. This quantity is familiar in the Swampland Programme literature because, for extremal BPS objects, it corresponds to the extremality factor that is crucial to properly formulate the WGC \cite{Palti:2017elp, Heidenreich:2019zkl}. In our case it will play a different role, namely as a measure of how much an EFT gauge subsector decouples from gravity, and therefore becomes an RFT. We show that for limits in which an RFT decouples gravitationally the charge-to-mass ratio of particles that are purely charged under the RFT gauge group -- computed as if they were BPS -- go to infinity, and the other way around. This relation between rigid limits and $\gamma$'s that diverge can be determined purely at the IR level, but the existence of actual BPS particles that correspond to such $\gamma$'s cannot, and that is where the scalar curvature enters. Gathering the results from \cite{Marchesano:2023thx,Marchesano:2024tod,CMP}, we find that whenever $\mathsf{R}$ diverges, there is  at least one BPS particle populating a quantised charge vector $\mathsf{q}$ such that $\gamma_\mathsf{q}$ blows up. Moreover, we find that $\mathsf{R}_{\rm div} \lesssim  \gamma_{\max}^2$, where $\gamma_{\max}$ is the $\gamma_\mathsf{q}$ with largest divergence. Together with the previous result, this provides further support to the proposal that curvature divergences imply the decoupling of an RFT from gravity, dubbed Curvature Criterion in \cite{Marchesano:2023thx}.

More precisely our results suggest that, along geodesics in the vector moduli space where the scalar curvature diverges, it shows an asymptotic behaviour of the form
\be
\mathsf{R}_{\rm div} \sim \left(\frac{\Lambda_{\rm wgc}}{\Lambda_g}\right)^2 ,
\label{Rdiv1}
\ee
with again $\Lambda_{\rm wgc} = g_{\rm rigid} \Mpl$  the WGC-predicted cut-off scale for the RFT sourcing the curvature divergence. The scale $\Lambda_g = g_{\rm rigid}^{-2} \Lambda_{\rm RFT}$, on the other hand, roughly captures the electrostatic energy of the weakly-coupled RFT, integrated up to its actual cut-off $\Lambda_{\rm RFT}$. In this sense, one can interpret the curvature divergence as a quotient between a gravitational UV scale and its field-theoretic counterpart. For a divergence to occur, these two scales must differ parametrically along the limit, which means that all states charged under the RFT go below the QG-predicted cut-off. This, in turn, implies a decoupling of the RFT from gravity, in agreement with our intuition. Finally, it follows that knowledge of $\mathsf{R}$ and $g_{\rm rigid}$ allows us to characterise $\Lambda_{\rm RFT}$, providing a further instance of the non-trivial interplay between IR and UV data.

\section{Moduli space curvature in type II Calabi--Yau compactifications}
Let us consider type II string theory compactified on a Calabi--Yau threefold. The Lagrangian that describes the vector multiplet sector of the resulting 4d ${\cal N}=2$ Effective field Theory (EFT) reads \cite{Ferrara:1988ff,Andrianopoli:1996cm,Lauria:2020rhc}
\begin{align}
\label{SVM}
S_{\rm 4d}^{\rm VM} & =  \frac{1}{2\kappa_{4}^2} \int_{\mathbb{R}^{1,3}} R * \mathbbm{1} - 2 g_{i\bar{j}} dz^i \wedge * d\bar{z}^{\bar{j}} \\ & +  \frac{1}{2} \int_{\mathbb{R}^{1,3}} \CI_{IJ} F^I \wedge * F^J + \CR_{IJ} F^I \wedge F^J\,  ,
\nonumber
\end{align}
where we follow similar conventions to \cite{Freedman_VanProeyen_2012}. Here, $z^i$ are the complex fields that parametrise the vector multiplet moduli space, and $g_{i\bar{j}}$ defines a special K\"ahler metric. For type IIB string theory compactified on a Calabi--Yau $Y_3$, 
the K\"ahler potential for this metric is globally defined in terms of its holomorphic three-form $\Omega$ as \cite{Strominger:1990pd}
\be
K = - \log \left(i\int_{Y_3} \Omega (z) \wedge \bar{\Omega} (\bar{z})  \right)\, .
\ee
The couplings in the Lagrangian \eqref{SVM} are easier to describe in terms of special coordinates. As argued in \cite{Craps:1997gp}, for each local patch in moduli space there is always a choice of symplectic three-cycle basis $(A^I, B_J)$ such that they exist. This basis moreover satisfies $A^I \cdot A^J = B_I \cdot B_J = 0$, as well as $A^I \cdot B_J = \delta^I_J$,  $I = (0,i) = 0, 1, \dots,  h^{2,1}(Y_3) = n_V$, with its Poincar\'e dual basis of harmonic three-forms $(\beta^I, \alpha_J)$ verifying $\int_Y \alpha_J \wedge \beta^I = \delta^I_J$. Upon expanding $\Omega$ on the latter one finds
\be
\Omega(z) = X^I (z) \alpha_I - \CF_J (z) \beta^J\, , 
\ee
with $\CF_I = \p_{X^I} \CF$. Here $\CF$ is a holomorphic, homogeneous degree-two function of the periods $X^I = \int_{A^I} \Omega$, dubbed the prepotential. One can  describe the coordinates in \eqref{SVM} as $z^i = X^i/X^0$, such that the moduli space metric reads
\be
g_{i\bar{j}} =  k_i k_{\bar{j}} + 2|X^0|^2 e^{K} \im \CF_{ij}\, ,
\label{IIBmetric}
\ee
where $\CF_{ij} = \p_{X^i} \p_{X^j} \CF$ and we have introduced
\be
k_i = -i e^K \int_{Y_3} \p_{z^i} \Omega \wedge \bar{\Omega}\, .
\label{monok}
\ee
Notice that the factors of $X^0$ cancel out, so that \eqref{IIBmetric} can be seen as a function of $(z^i, \bar{z}^{\bar{j}})$. Finally, one has
\be
\mathcal{N}_{IJ} \equiv \CR_{IJ} + i \CI_{IJ} = \bar{\CF}_{IJ} + 2i \frac{\im \CF_{IK}\im \CF_{JL} X^K X^L}{\im \CF_{MN} X^M X^N }\, ,
\ee
where $\CR_{IJ}, \CI_{IJ} \in \mathbb{R}$ should be again understood as functions of $(z^i, \bar{z}^{\bar{j}})$, which can be thus related to the metric in field space as follows \cite{Strominger:1990pd, Andrianopoli:1996cm,Craps:1997gp,Freed:1997dp}
\be
g_{i\bar{j}} = -2  f^M_i \CI_{ML} \bar{f}^L_{\bar{j}} = 2 f^M_i \im \CF_{ML} \bar{f}^L_{\bar{j}}\, ,
\label{identity}
\ee
with $f_i^L = -e^{K/2} \int_{A^L} (\p_{z^i} \Omega + k_i \Omega)  = e^{K/2} (X^0  \delta_i^L + k_i X^L)$. 

In this framework, one can easily compute the moduli space scalar curvature, yielding (in Planck units) \cite{Strominger:1990pd,Andrianopoli:1996cm}
\be
\mathsf{R} = - 2 n_V(n_V + 1) + 2 |X^0|^6 e^{2K} g^{i\bar{j}}g^{k\bar{l}}g^{m\bar{n}}  \CF_{ikm} \bar{\CF}_{\bar{j}\bar{l}\bar{n}}\, ,
\label{IIBscalar}
\ee
where $\CF_{ijk} \equiv \p_{X^i} \p_{X^j} \p_{X^k} \CF$. Clearly $\mathsf{R} (z)$ is bounded from below, and the only way in which one can generate a divergence is when some term(s) within the sum in the right hand side of \eqref{IIBscalar} blow up. It was moreover found in \cite{Marchesano:2023thx} that, for infinite distance limits pertaining to the large complex structure regime, such behaviour is related to certain gauge interactions decoupling from gravity asymptotically. It is our aim here to refine and extend this picture within the present, more general setup. 

\section{How to get a rigid theory}
The vector-multiplet Lagrangian of a rigid 4d $\CN=2$ field theory is also specified by some holomorphic function $\mathscr{F}(z^\mu)$ \cite{Grimm:1977xp,Breitenlohner:1981sm}. This time, the gauge kinetic function is simply given by $\mathscr{F}_{\mu\nu} \equiv \p_{z^\mu} \p_{z^\nu} \mathscr{F}$, whereas the K\"ahler potential reads $\mathscr{K} = 2\im \left[ \bar{z}^{\bar{\mu}} \p_{z^\mu} \mathscr{F} \right]$, from which the rigid relation $G_{\mu\bar{\nu}} = 2 \im \mathscr{F}_{\mu\nu}$ follows immediately. Several schemes have been discussed in the literature in order to recover a rigid field theory from its local counterpart, particularly so in the context of type II Calabi--Yau compactifications \cite{Seiberg:1994rs,Kachru:1995fv,Klemm:1996bj,Andrianopoli:1996cm,Katz:1996fh,Billo:1998yr,Gunara:2013rca,Alexandrov:2017mgi}. In the following, we offer a different, general description of how to obtain rigid field theories from limits lying purely within the vector multiplet moduli space, that is more suitable for our purposes. 

The rigid theory that one recovers from a supergravity EFT inherits its prepotential from the latter, but only a subset $\{ z^\mu \} \subset \{z^i \}$ can be considered as \emph{dynamical} fields in the RFT. One may indeed identify such fields by looking at \eqref{IIBmetric}, where $\CF_{\mu\nu}$ reduces to the gauge kinetic function $\mathscr{F}_{\mu\nu}$ of the rigid theory. It then follows that the field directions belonging to the RFT must satisfy $g_{\mu\bar{\nu}} \sim 2\im \CF_{\mu\nu}$, or more precisely require that
\be
\frac12 e^{-K} |X^0|^{-2} k_\mu k_{\bar{\nu}} \ll \im \CF_{\mu\nu} \sim g_{\rm rigid}^{-2}\, .
\label{cond}
\ee 

To obtain a physical interpretation of this condition, let us consider a D3-brane wrapping a special Lagrangian three-cycle in the integer homology class $[\Pi_3] = q_I [A^I] - p^I [B_I]$. Upon dimensional reduction, we find a 4d particle sourcing the field strengths $(F^I, G_I)$ -- with $G_I \equiv \delta \CL / \delta F^I$ the magnetic dual of $F^I$ -- with charges $\mathsf{q}^{\rm T} =  (p^I, q_I)$ \cite{Ceresole:1995ca}. The particle mass is given by its central charge $Z_\mathsf{q}$
\be
m_\mathsf{q}  =  |Z_\mathsf{q}|\, \Mpl\, , \quad Z_\mathsf{q}  = e^{K/2} \left(q_I X^I - p^I \CF_I\right)\, , 
\label{massD3}
\ee
with $\Mpl = \sqrt{8\pi}/\kappa_4$, whereas its squared physical charge can be computed as
\be
\CQ^2_\mathsf{q}  \equiv - \oh \tilde{\mathsf{q}}^{\rm T} \begin{pmatrix} \CI  \\  & \CI^{-1} \end{pmatrix} \tilde{\mathsf{q}} \, ,
\label{chargeD3}
\ee
where
\be
\tilde{\mathsf{q}} = \begin{pmatrix} \id  \\ - \CR & \id \end{pmatrix} \mathsf{q}  =  \begin{pmatrix} p^I \\ q_I - \CR_{IJ}p^J \end{pmatrix}\, .
\label{vecQ}
\ee
The rigid version of the physical charge, which we dub $\CQ_{\rm rigid}^2$, is obtained from  \eqref{chargeD3} upon replacing $\mathcal{N}_{IJ} \to \bar{\CF}_{IJ}$.

Let us consider a charge vector $\mathsf{q}$ such that \footnote{This relation can be satisfied exactly for certain choices of moduli fields, or to arbitrary accuracy for generic ones.}
\be
p^I = 0\ \text{for}\ I \notin \{\mu\}\, , \qquad q_I = \re \CF_{I\mu} p^\mu\, .
\label{chcharges}
\ee
Then its rigid and central charges read
\be
\CQ_{\mathsf{q},{\rm rigid}}^2 = \oh \im \CF_{\mu\nu} p^\mu p^\nu\, , \quad
Z_\mathsf{q} =  -\frac{i}{2} \frac{e^{-K/2}}{\bar{X}^0}  k_{\bar{\mu}}p^\mu \, ,
\label{QZrig}
\ee
where we have used that
\be
 k_{\bar{\nu}} = 2 \bar{X}^0 e^{K} \im \CF_{\nu J} X^J\, .
 \label{knu}
\ee
Hence, if this particle is BPS one can interpret \eqref{cond} as
\be
m_\mathsf{q}^2 \ll \CQ_{\mathsf{q},{\rm rigid}}^2 \Mpl^2 \implies 
\gamma_\mathsf{q}^2 > \gamma^2_{\mathsf{q},{\rm rigid}} \frac{\Mpl^2}{m_*^2} \gg 1 \, ,
\ee
where $\gamma_\mathsf{q}$ is the charge-to-mass ratio of a BPS particle with charge $\mathsf{q}$ in the supergravity theory \footnote{For extremal BPS states, $\gamma$  corresponds to the extremality factor. Given the results of \cite{Gendler:2020dfp,Alim:2021vhs}, one does not expect RFT-charged particles to be extremal in general, so we henceforth refer to $\gamma$ as charge-to-mass ratio. It would nevertheless be interesting to compare our findings  with the divergences for $\gamma$ recently found in \cite{FierroCota:2023bsp}.}, whilst
\be
\gamma_{\mathsf{q},{\rm rigid}}^2 \equiv \frac{\CQ_{\mathsf{q},{\rm rigid}}^2}{|Z_{\mathsf{q},{\rm rigid}}|^2}\, , \  Z_{\mathsf{q},{\rm rigid}} \equiv (q_I X^I - p^I \CF_I)/X^0\, ,
\label{gammastar}
\ee
denotes its rigid analogue. Here $|Z_{\mathsf{q},{\rm rigid}}| = m/m_*$  measures the particle mass in units of the reference scale
\be
m_* = m_{\rm D3}(A^0) = e^{K/2} |X^0| \Mpl\, .
\label{mstar}
\ee
Notice that it is by expressing the kinetic terms in units of this scale that one recovers a rigid Lagrangian, since
\be
\hat{g}_{i\bar{j}} \equiv |X^0|^{-2} e^{-K} {g}_{i\bar{j}} \stackrel{\eqref{cond}}{\longrightarrow} 2 \im \CF_{\mu\nu} = G_{\mu\bar{\nu}}\, .
\label{cond2}
\ee
Simultaneously, we have that $\CI_{\mu\nu} \to  -\im \CF_{\mu\nu}$ for the corresponding $U(1)$ factors.

To sum up, the condition \eqref{cond} requires the charge-to-mass ratio of a BPS particle charged under the rigid theory to diverge along the rigid limit:
\be
\gamma^2_\mathsf{q}  \equiv \frac{\CQ^2_\mathsf{q} \Mpl^2}{m^2_\mathsf{q} } \to \infty \, .
\label{extrema}
\ee

Reciprocally, $U(1)$ factors whose charged states satisfy \eqref{extrema} are candidates for a 4d $\CN=2$ RFT decoupling from gravity. Given a BPS state with charge $\mathsf{q}$, one can express its charge-to-mass ratio as \cite{Ceresole:1995ca,Palti:2017elp}
\be
\gamma_\mathsf{q}^2  = 1 + 4 g^{i\bar{j}} \p_i \log \left( \frac{m_\mathsf{q}}{\Mpl} \right)\, \bar{\p}_{\bar{j}} \log \left( \frac{m_\mathsf{q}}{\Mpl} \right) ,
\label{gamma}
\ee
where the mass should be regarded as a function of $(z^i, \bar{z}^{\bar{j}})$, as per \eqref{massD3}. Additionally, one has \cite{Ceresole:1995ca}
\be
\gamma^2_{\mathsf{q},{\rm rigid}} \frac{\Mpl^2}{m_*^2} = - 1 + 4 g^{i\bar{j}} \p_i \log \left( \frac{m_\mathsf{q}}{\Mpl} \right)\, \bar{\p}_{\bar{j}} \log \left( \frac{m_\mathsf{q}}{\Mpl} \right) ,
\label{gammarigidid}
\ee
%
from where one obtains $\gamma_\mathsf{q}^2 = 2 + \gamma^2_{\mathsf{q}, {\rm rigid}} (\Mpl^2/m_*^2)$. 
Hence, in the limit $\gamma_\mathsf{q}^2 \to \infty$ we recover the (asymptotic) relation $\gamma_\mathsf{q}^2 \sim \gamma^2_{\mathsf{q}, {\rm rigid}} (\Mpl^2/m_*^2)$, which signals that we are approaching a rigid theory regime. To identify an actual RFT subsector, we will in addition require that the kinetic mixing between the RFT directions $\{z^\mu\}$ and the rest vanishes sufficiently fast, which can be rephrased as the condition that $g^{\mu\bar{\nu}}$ reduces to the inverse of $2|X^0|^2 e^K \im \cF_{\mu\nu}$ along the limit.  This extra condition is standard in field theory, and it is satisfied for each of the setups discussed in the following subsections.

Note that, neither to define $\gamma_\mathsf{q}$ nor the RFT condition \eqref{extrema} it is necessary for the charge $\mathsf{q}$ to host a stable BPS particle. Nevertheless, in  the limits that we will analyse later on, in which the moduli space curvature diverges, such RFT charged particles do exist. In that case, the relation between a diverging charge-to-mass ratio $\gamma$ and an RFT limit can be understood very intuitively, because in both cases gauge interactions are parametrically stronger than gravity, and the no-force condition between mutually BPS particles is exclusively attained due to the exchange of scalar and spin-1 fields \cite{Palti:2017elp,Heidenreich:2019zkl}. 

Roughly speaking, the two main scenarios where \eqref{extrema} can occur is when we take {\it a)} some $U(1)$'s to strong coupling while keeping their electrically charged BPS particles with finite masses in Planck units or {\it b)} some electric masses to vanish in Planck units while keeping their gauge couplings constant or slowly decreasing. In some instances, the first option can be translated into the second one via electric-magnetic duality \cite{Seiberg:1994rs}. When not, it is likely because we are dealing with a strong coupling singularity in the interior of moduli space. For this reason we will mostly have in mind option {\it b)}, that is both realised at finite-distance conifold-like degenerations \cite{Strominger:1995cz} and at infinite-distance asymptotic limits \cite{Marchesano:2023thx,Marchesano:2024tod,CMP} in vector multiplet moduli space. 

For infinite-distance singularities,  refs. \cite{Grimm:2018ohb,Grimm:2018cpv} have shown that there are certain gauge couplings that vanish along the limit. In addition, the charge-to-mass ratio $\gamma$ of the extremal BPS particles charged under them are asymptotically constant \cite{Gendler:2020dfp,Bastian:2020egp}. This happens for instance for the field direction $t \to \infty$ along the asymptotic trajectory, whose kinetic terms can be well approximated by a K\"ahler potential of the form $K \simeq - n \log t$, for some $n \in \mathbb{N}$ \cite{Grimm:2018cpv}. Hence, upon considering a particle electrically charged under this direction one finds that both its physical charge and mass in Planck units scale as $e^{K/2} t \sim t^{1-n/2}$, leading to an asymptotically  finite $\gamma$.

It turns out that for this class of limits there is a natural choice for the reference scale \eqref{mstar}, namely the characteristic mass of the descending tower of states predicted by the Distance Conjecture (SDC) \cite{Ooguri:2006in}. The results of \cite{Grimm:2018ohb,Grimm:2018cpv}  show that this scale may be identified with the three-cycle generated by the monodromy at the infinite-distance singularity, which is naturally an electric cycle in the symplectic basis. As a result,  it can be taken to be $A^0$ without loss of generality, so in this frame we have that $ \Mpl^2/ m_*^2 = e^{-K}/|X^0|^2 \simeq t^n$ diverges along the limit. This, in turn, induces a divergence for $\gamma_{\mathsf{q}}$, unless it is compensated by  a $\gamma_{\mathsf{q},{\rm rigid}}$ decreasing  sufficiently fast, as precisely happens in the case just discussed. 

In general, along RFT limits there will be  $\gamma_{\mathsf{q}}$'s with different degrees of divergence. Out of the different RFT subsectors with diverging  $\gamma_{\mathsf{q}}$'s and vanishing mixing, we will identify the dominant RFT sector with those field directions where the condition \eqref{cond} is attained faster, which are associated to the set of $\gamma_{\mathsf{q}}$'s which exhibit the largest divergence along the limit. The relation between RFT field directions and the spectrum of divergences for $\gamma_{\mathsf{q}}$ along the lattice of charges is in general non-trivial, but it simplifies if the RFT directions $\{z^\mu\}$ also contain the parametrically largest gauge couplings. In that case, the set of charges $\mathsf{q}$ of the form \eqref{chcharges} that have maximal divergence $\gamma_{\mathsf{q}} \sim \gamma_{\rm max}$ form a sublattice, from whose generators one can determine the RFT field directions. In particular, the number $N_\gamma$ of linearly independent charges \eqref{chcharges} such that $\gamma_{\mathsf{q}} \sim \gamma_{\rm max}$ can then be identified with the rank of the dominant RFT gauge group  
\be
r_{\rm rigid} = N_{\gamma}\, .
\label{rrigdef}
\ee

Let us now consider an infinite distance limit that falls into scenario {\it b)}. If the RFT condition \eqref{extrema} is satisfied, this setup must display a parametric hierarchy in the $\gamma_{\mathsf{q}}$'s, whose values may either asymptote to a constant or present different degrees of divergence. In general, one expects this structure to be sourced by a hierarchy in the masses of BPS charged particles together with a hierarchy of gauge couplings which, per \eqref{identity}, also imply a hierarchy of metric eigenvalues. In this setup, typically the largest charge-to-mass ratios $\gamma_{\mathsf{q}} \sim \gamma_{\max}$ will correspond to the smallest metric eigenvalues and the largest gauge couplings, which can be identified with the RFT field directions. An exception to this rule is the direction along the asymptotic trajectory, whose gauge coupling is not suppressed along certain infinite distance limits \cite{Marchesano:2023thx}. However, as already discussed the charge-to-mass ratio associated to this field asymptotes to a constant, and therefore it must not be considered as part of the RFT.

\section{Rigid limits and curvature}

Let us consider an infinite-distance RFT limit which in addition displays a hierarchy of metric eigenvalues. By the consideration above, one expects that $g_{\mu\bar{\nu}}$ are much smaller than the average metric entries, with $\{z^\mu\}$ the fields that belong to the RFT \footnote{As pointed out already, this may also apply for the field associated to the asymptotic trajectory, but in practice one can check that this direction does not contribute to the curvature divergence \cite{Marchesano:2023thx,Marchesano:2024tod,CMP}.}. In that case, the right-hand side of \eqref{IIBscalar} is dominated by the term
\be
\mathsf{R}_{\rm div} \sim e^{-K} 2 \hat{g}^{\mu\bar{\nu}} \hat{g}^{\rho\bar{\sigma}} \hat{g}^{\tau\bar{\eta}}  \CF_{\mu\rho\tau} \bar{\CF}_{\bar{\nu}\bar{\sigma}\bar{\eta}} 
 \sim \frac{\Mpl^2}{m_*^2} R_{\rm rigid}\, ,
\label{Rdiv2}
\ee
where
\be
\mathsf{R}_{\rm rigid} = 2 G^{\mu\bar{\nu}} G^{\rho\bar{\sigma}} G^{\tau\bar{\eta}}  \mathscr{F}_{\mu\rho\tau} \bar{\mathscr{F}}_{\bar{\nu}\bar{\sigma}\bar{\eta}}\, , 
\label{Rrigid}
\ee
with $\mathscr{F}_{\mu \nu \rho} = X^0 \CF_{\mu \nu \rho}$ and $G_{\mu\bar{\nu}} =2 \im \mathscr{F}_{\mu\nu}$ the corresponding rigid quantities. Note that \eqref{Rdiv2} is precisely the expression used in \cite{Marchesano:2023thx,Marchesano:2024tod} to detect and classify curvature divergences along infinite-distance limits. Remarkably, the same relation is obtained for finite-distance, weak-coupling singularities where $m_{\rm el}/ m_* \to 0$ \cite{CMP}. In this case, $e^{-K}/|X^0|^2 = \Mpl^2/ m_*^2$ is bounded from above, and $\gamma^2$ diverges purely because $\gamma^2_{\rm rigid}$ does. In terms of the curvature, the divergence can be traced back to one or several poles in $\CF_{\mu\rho\tau}$. This selects the field directions $z^\mu$ that control the mass of the light charged states within the RFT, and again provides a structure for the curvature divergence of the form \eqref{Rdiv2}.

If instead we assume that we have a curvature divergence, then clearly it can be written in the form \eqref{Rdiv2} for some choice of field subset $\{z^\mu\}$. In the weakly-coupled regimes that we are presently considering, the source of the divergence may either come from $\Mpl^2/ m_*^2$ or $\mathscr{F}_{\mu \nu \rho}$, which again takes us to the infinite/finite distance scenarios as discussed above, or  to a combination of both \footnote{Examples of these are the Seiberg-Witten points discussed in the text, or the so-called coni-LCS singularities \cite{Demirtas:2020ffz,Alvarez-Garcia:2020pxd,Bastian:2021eom}, which can also be shown to obey both \eqref{Rdiv1} and \eqref{bound} \cite{CMP}.}. As a result, one is always able to select a RFT that reproduces the divergence via \eqref{Rdiv2}. This is, essentially, the content of the Curvature Criterion put forward in \cite{Marchesano:2023thx}. 

Given the relation between rigid field theory limits, divergences for charge-to-mass ratios and the moduli space scalar curvature, it is thus natural to compare the last two. In the following, we will argue that divergences of $\gamma$ bound from above the ones appearing in $\mathsf{R}$. More precisely, we show that the following bound 
\be
\mathsf{R}_{\rm div} \lesssim r_{\rm rigid} \, \gamma_{\rm max}^2
\label{bound}
\ee
holds in a wide class of rigid field theory limits, with $\gamma_{\rm max}$ defined as in the previous section. Note that, in terms of RFT quantities, proving this inequality amounts to show 
\be
\mathsf{R}_{\rm rigid} \lesssim r_{\rm rigid} \, \gamma_{\rm rigid}^2 \, ,
\label{boundrigid}
\ee
with $\mathsf{R}_{\rm rigid}$ defined as above and $\gamma_{\rm rigid}$ the rigid version of $\gamma_{\rm max}$.  Note that one can express rigid $\gamma$'s as \footnote{The identity \eqref{gammarigid} can be deduced from the no-force condition $\CQ_{\rm rigid}^2 = G^{\mu \bar{\nu}} \partial_{\mu} Z_{\rm rig} \bar{\partial}_{\bar{\nu}} \bar{Z}_{\rm rig}$, which holds for BPS states in 4d $\CN=2$ rigid field theories.}
\be
\gamma_{\mathsf{q},{\rm rigid}}^2 = 4 G^{\mu \bar{\nu}} \p_{\mu} \log \left( \frac{m_\mathsf{q}}{m_*} \right)\, \bar{\p}_{\bar{\nu}} \log \left( \frac{m_\mathsf{q}}{m_*} \right)\, .
\label{gammarigid}
\ee
Notice that condition \eqref{bound}, together with the relation between $\gamma$ divergences and RFT limits, imply the Curvature Criterion in the present setup. In what follows, we will illustrate how this bound is satisfied for a large class of limits within scenario {\it b)}, which essentially avoids finite-distance, strong coupling singularities of the superconformal type (see e.g. \cite{Akhond:2021xio} and references therein). Instead of the type IIB framework employed up to now, we will place our discussion in the mirror dual type IIA Calabi--Yau compactifications, since this is the setup in which the analysis of \cite{Marchesano:2023thx,Marchesano:2024tod,CMP} is carried out.

\section{Large volume divergences}
In the large volume regime of type IIA compactified on a Calabi--Yau threefold $X_3$, the full prepotential of the 4d $\CN =2$ supergravity EFT reads
\begin{align}\nonumber
 {\cal F}/(X^0)^2 = & -\frac{1}{6} \CK_{ijk}z^i z^j z^k + \oh K_{ij}^{(1)} z^i z^j + K_{i}^{(2)} z^i + \frac{i}{2} K^{(3)}\\ & + (2\pi i)^{-3} \sum_{\boldsymbol{k} > \boldsymbol{0}} n_{\boldsymbol{k}}^{(0)} {\rm Li}_3 \left( e^{2\pi ik_i z^i} \right) ,
\label{fullFLV}
\end{align}
where we have written $\CF$ in terms of the affine coordinates $z^i = X^i/X^0$. In this regime, the latter can be identified with the complexified K\"ahler coordinates:
\be
J_c = B + i J  = (b^i + i t^i) \omega_i = z^i \omega_i\, , 
\ee
where $i =1, \dots, h^{1,1}(X_3) = n_V$ and $\{\omega_i\}$ is dual to a basis of Nef divisors of $X_3$. The different coefficients in \eqref{fullFLV} depend on the topology of the threefold:  $\CK_{abc} = \int_{X_3} \omega_a\wedge \omega_b \wedge \omega_c$ are the triple intersection numbers, whereas $n_{\boldsymbol{k}}^{(0)}$ is the genus-zero Gopakumar-Vafa (GV) invariant of the curve class $k_i {\cal C}^i$, with ${\cal C}^i$ such that $ \int_{{\cal C}^i} \omega_j = \delta^i_{j}$. Li$_3 (y)$ stands for the 3rd polylogarithmic function and $K^{(i)}$ $i=1,2,3$ are further topological data of $X_3$, see e.g. \cite{Marchesano:2022axe} for their precise definition. 

The asymptotic behaviour of the moduli space scalar curvature along infinite-distance limits within the regime $t^i \gg 1$ has been analysed in detail \cite{Marchesano:2023thx,Marchesano:2024tod}. The kind of trajectories considered are of the form
\be
\boldsymbol{t} = \boldsymbol{e}_0 \phi  +  \boldsymbol{e}_1 \phi^{\delta_1} +  \boldsymbol{e}_2 \phi^{\delta_2} + \dots \, ,
\label{growth}
\ee
with an appropriate rescaling of the 10d dilaton. Here $\boldsymbol{t}$ is the vector of saxionic K\"ahler coordinates, and $\boldsymbol{e}_\alpha$ are vectors of non-negative integers representing Nef divisors $\CD_{\boldsymbol{e}_\alpha} = e_\alpha^i [\omega_i]$, such that their Cartesian product satisfies $\boldsymbol{e}_\alpha \cdot \boldsymbol{e}_\beta = 0$, $\forall \alpha \neq \beta$. Additionally $1> \delta_1 > \delta_2 > \dots > 0$ and $\phi \to \infty$ is the parameter that describes the  trajectory. Upon appropriate reparametrisations, these limits are expected to asymptotically approximate all infinite-distance geodesics in this moduli space region \cite{Lanza:2021udy}. 

Along any of the aforementioned limits the lightest tower of massive states is given by the one of D0-branes, which set the scale $m_*$ via the replacement $m_{\rm D3} \to m_{\rm D0}$ in \eqref{mstar}, as expected from mirror symmetry \cite{Strominger:1996it}. Additional towers of D2, D4 and NS5-branes may combine with the D0's  multiplicatively \cite{Castellano:2021mmx}, in order to lead to F-theory decompactifications and/or emergent string limits. Regardless of this casuistic, it is the D0-brane/SDC tower scale that sets the rescaling \eqref{cond2} which takes us from the supergravity to the RFT metric. In this type IIA setup, the conversion factor is given by
\be
\frac{\Mpl^2}{m_*^2} = 8 \CV_{X_3} \, , 
\ee
with $\CV_{X_3} = {\rm Vol}(X_3)$, which always grows to infinity along infinite-distance type IIA large volume trajectories \cite{Corvilain:2018lgw}. This indeed matches with the approximate expression for the K\"ahler potential in this regime
\be
K \simeq - \log (8 \CV_{X_3}) = - \log \left(\frac{4}{3} \CK_{ijk} t^i t^j t^k\right)\, , 
\label{KIIA}
\ee
which discards the contribution of the curvature correction $K^{(3)}$ and the exponential terms generated by world-sheet instantons. Even when neglecting these corrections, the metric derived from \eqref{KIIA} is non-degenerate, and suffices to determine when an RFT limit with a hierarchy of metric eigenvalues is attained along trajectories of the form \eqref{growth}. Following \cite{Marchesano:2023thx}, let us consider the matrix
\be
{\bf K}_{ij} \equiv  \CK_{ijk} e^k_0\, ,
\label{matrixK}
\ee
and compute its rank $r_{\bf K}$. A non-maximal rank implies that there is one or more divisors $\CD_\mu$ that do not intersect the divisor $\CD_{\boldsymbol{e}_0}$ which defines the leading term in \eqref{growth}. Let us assume that such divisors are in addition effective. Then, for the choice  $\delta_{\alpha > 0} =0$, they remain of constant volume (in string units) along the limit, and so do their mutual intersections. A D2-brane wrapping the curve $\CD_\mu \cdot \CD_\nu$ will give rise to a particle with constant mass in units of $m_*$, and whose rigid charge $\CQ_{\rm rigid}$ is bounded from below, since by construction $\im \CF_{\mu\nu}$ has a constant polynomial piece. It follows that we have diverging charge-to-mass ratios $\gamma$ for such particles, and hence an RFT-decoupling limit. Switching on $\delta_{\alpha > 0} > 0$ will change some of the above statements, but since these correspond to sub-leading effects it will not modify the fact that $\gamma$ diverges and that a gauge sector decouples from gravity. One can moreover check that the rank of such an RFT is given by
\be
r_{\rm rigid} = n_V - r_{\bf K} - \lambda\, , 
\ee
with $\lambda =0$ for decompactification limits and $\lambda = 1$ for emergent string limits. This is because for emergent string limits the divisor $\CD_{\boldsymbol{e}_0}$ belongs to  $\ker {\bf K}$ but does not intersect itself nor any other element of  $\ker {\bf K}$. As a result it does not imply a diverging $\gamma$, and indeed the corresponding $U(1)$ generator should not be considered as part of the RFT, as stated previously. Finally, an important property of these limits is that to obtain the rigid prepotential $\CF_{\rm rigid}$  one only needs to select those terms of \eqref{fullFLV} that depend on the RFT fields $\{z^\mu\}$.

Having specified the RFT in terms of $\ker {\bf K}$, one obtains a scalar curvature divergence of the form \eqref{Rdiv2} \footnote{When $\ker {\bf K}$ is trivial, one can also have a set of diverging $\gamma$'s and an RFT-decoupling limit. However, in this case the RFT gauge couplings tend to zero quite fast, and as a result the asymptotic behaviour of the metric eigenvalues is such that no curvature divergence is generated. At most, one may have a term of the form \eqref{Rdiv2} that contributes as a constant to the asymptotic curvature, rendering it in some cases positive \cite{Marchesano:2023thx}.}. Then, using the results of \cite{Marchesano:2023thx,Marchesano:2024tod}, one can distinguish between three kinds of limits:

\begin{itemize}

 \item[{\it i)}] The curvature diverges at the classical level, namely for the approximate metric that one derives from \eqref{KIIA}. This happens whenever $\CK_{\mu\nu\rho} \neq 0$ among the divisors $\{ \CD_\mu\}$ defining the RFT. Then $\CF_{\rm rigid}$ has a cubic term in its polynomial piece. 

 \item[{\it ii)}] The curvature only diverges if one takes into account the exponential terms in \eqref{fullFLV}. In this case $\CK_{\mu\nu\rho} = 0$ and the RFT metric is flat unless world-sheet instanton corrections are taken into account. 

 \item[{\it iii)}] The rigid metric is flat, and there is no curvature divergence. This occurs either when the RFT displays an enhanced $\CN=4$ supersymmetry, or when $\CK_{\mu\nu\rho} = 0$ and none of the divisors $\{ \CD_\mu\}$ is effective.

\end{itemize}

For the first two scenarios, the asymptotic behaviour of the curvature is described by \eqref{Rdiv2}. In particular, in the first case $\CF_{\mu\rho\tau}$ can be approximated by the real constants $\CK_{\mu\rho\tau}$, and we have that $\mathsf{R}_{\rm rigid} \lesssim r_{\rm rigid} \max \{g_{\rm rigid}\}^6$, where $\max \{g_{\rm rigid}\}$ is the largest of the gauge couplings within the RFT. When there is a single gauge coupling -- or rather all of them are of the same order -- the equality is saturated, and one obtains
\be
\mathsf{R}_{\rm div} \sim r_{\rm rigid}^3 \left(\frac{\Lambda_{\rm wgc}}{\Lambda_g}\right)^2 .
\label{RdivLV}
\ee
Here $\Lambda_{\rm wgc} \equiv g_{\rm rigid} \Mpl$ is the maximal RFT cut-off scale predicted by the WGC. Also $\Lambda_g = r_{\rm rigid} g_{\rm rigid}^{-2} \Lambda_{\rm RFT}$ is the  electrostatic energy of the RFT integrated up to its actual cut-off $\Lambda_{\rm RFT}$, which here is given by the Distance Conjecture scale $m_*$. When the RFT gauge couplings differ one has $\Lambda_{\rm wgc} = \min \{g_{\rm rigid}\} \Mpl$ and $\Lambda_g \lesssim r_{\rm rigid} \min \{g_{\rm rigid}\}^{-2} \Lambda_{\rm RFT}$. However, since by assumption all the charge-to-mass ratios of this sector scale like $\gamma_{\rm max}$ along the limit,  the couplings do the same, and so \eqref{RdivLV} still captures the scalar curvature behaviour. Indeed, looking at the charge-to-mass ratio \eqref{gammarigid} one has that for electrically charged particles $m \simeq |z^\mu| \simeq t^\mu \simeq \im \CF_{\mu\mu} \simeq g_{\rm rigid}^{-2}$, and so one finds
\be
\gamma_{\rm rigid}^2 \simeq  g_{\rm rigid}^6\, .
\ee
Notice that that the inequality \eqref{boundrigid} is also satisfied. 

If the curvature divergence is sourced by world-sheet instanton corrections, one has $\CF_{\mu\rho\tau} \sim  {\cal O}(e^{-2\pi t^\mu})$, while still the RFT gauge couplings can be approximated by a K\"ahler modulus $g_{\rm rigid}^{-2}\simeq \im \CF_{\mu\nu} \simeq t^E$ which is non-dynamical from the viewpoint of the RFT. Due to this suppression factor we have $\mathsf{R}_{\rm rigid} \lesssim r_{\rm rigid} g_{\rm rigid}^{6} {\cal O}(e^{-4\pi t^\mu})$ and one finds the inequality
\be
\mathsf{R}_{\rm div} \lesssim r_{\rm rigid}^3 \left(\frac{\Lambda_{\rm wgc}}{\Lambda_g}\right)^2 .
\label{RdivLVineq}
\ee
Note however that for trajectories of the form \eqref{growth} $\CF_{\mu\rho\tau}$ can either be considered as a constant, so that \eqref{RdivLV} is still valid, or vanishes exponentially fast in $\phi$, removing the curvature divergence. This suggests that \eqref{RdivLV} should still hold along infinite-distance geodesic trajectories. Finally, the charge-to-mass ratio \eqref{gammarigid} can be approximated by
\be
\gamma_{\rm rigid}^2 \simeq  g_{\rm rigid}^2 (t^\mu)^{-2} , 
\ee
and so \eqref{boundrigid} is again satisfied.

Remarkably, for all limits of the form \eqref{growth} where the curvature diverges, the scale $\Lambda_g$ coincides with the scale $\Lambda_{\rm ch}$ defined in \cite{Marchesano:2024tod}. This is the scale where a tower of states with arbitrary charges under the RFT gauge group appears, either from bound states of D2-branes or from NS5-branes wrapping ${\cal D}_\mu$. It does not, however, coincide with the magnetic monopole masses, which are made up from D4-branes wrapping ${\cal D}_\mu$ and lie generically at a higher scale.

\section{Divergences at Seiberg--Witten points} 
In the type IIA frame, Seiberg--Witten (SW) limits can be engineered by considering a Calabi--Yau $X_3$ exhibiting some K3 fibration over a $\mathbb{P}^1$ base, and as such they can be seen as emergent heterotic string limits \cite{Lee:2019wij}. Since the quantum volume of the fibre cannot be shrunk beyond stringy lengths, these limits proceed by growing the size of the $\mathbb{P}^1$ together with the (quantum-corrected) volume of $X_3$. In practice, this means that D0-brane bound states become again part of the leading tower predicted by the Distance Conjecture and that, just like in the large volume case, we can identify the D0 with the D3-brane wrapping the $A^0$-cycle in the type IIB frame. As a consequence, $m_*$ can again be taken to be the D0-brane scale and $\Mpl^2/m_*^2 = e^{-K}|X^0|^{-2}$ blows up along the limit. 

Unlike in the large volume regime, the full supergravity prepotential is in general hard to compute near SW points \cite{Kachru:1995fv,CMP}. Nevertheless, one can make use of asymptotic Hodge theory techniques (see e.g. \cite{vandeHeisteeg:2022gsp} and references therein) to describe how hierarchies between different moduli space metric eigenvalues are generated. If one expresses the periods of $\Omega(z)$ using a set of coordinates which are singled out by the Nilpotent Orbit Theorem \cite{Schmid:1973cdo}, one finds that they split into a polynomial and an exponentially suppressed piece, and the same applies to $e^{-K}$ \cite{Grimm:2018ohb}. Near SW points, it is often the case where the polynomial piece of $e^{-K}$ alone leads to a degenerate metric $g_{\rm pol}$, such that one needs the presence of certain additional exponential terms, dubbed \emph{metric essential instantons}, so as to render it well-behaved \cite{Bastian:2021hpc}. Such instanton-like terms signal the presence of exponentially suppressed metric eigenvalues, which are natural candidates to generate a curvature divergence via the presence of some rigid field theory. Relatedly, each metric essential instanton corresponds to a linearly independent vector $\mathsf{q}$ of the form \eqref{chcharges}, with a diverging $\gamma_{\mathsf{q}}$. Along certain directions we have that $\gamma_{\mathsf{q}} \sim \gamma_{\rm max}$ for all of them \cite{CMP}, and so applying \eqref{rrigdef} one is led to 
\be
r_{\rm rigid} = n_V - {\rm rank} (g_{\rm pol})\, . 
\ee
%

While in explicit string theory constructions it is possible to directly check \eqref{bound}, in order to get an overall idea on how this bound works one can instead analyse \eqref{boundrigid} in SW theory. Let us consider for instance a $SU(2)$ gauge prepotential around the asymptotically free point \cite{Seiberg:1988ur, Seiberg:1994rs}
\begin{align}\label{eq:weakcouplingprepotential}
   \mathcal{F}_{\rm sw} (a)\,  =\, \frac{i a^2}{2\pi}  \left[ \log \left( \frac{a}{\Lambda_{\rm sw}}\right)^2 -3 - \oh \sum_{k=1}^{\infty} c_k \left( \frac{\Lambda_{\rm sw}}{a}\right)^{2k}\right] 
\end{align}
where $a$ denotes the rigid field controlling the mass of the $W$-boson and $\Lambda_{\rm sw} < |a|$ is the scale dynamically generated at one-loop, both measured in units of $m_*$. Here, one finds
\be\label{eq:curv&gaugecouplSU(2)}
\mathsf{R}_{\rm sw} \sim \frac{g_{\rm rigid}^6}{8 \pi^4 |a|^2}\, , \quad \text{with} \quad g_{\rm rigid}^{-2} = \frac{1}{2\pi^2} \log \left| \frac{a}{\Lambda_{\rm sw}} \right|\, , 
\ee
where $g_{\rm rigid}$ is the gauge coupling measured at the IR scale $\mu = |a|m_*$. Since $\gamma_{\rm sw} \sim g_{\rm rigid}/|a|$, and by assumption $g_{\rm rigid} \ll 1$, both \eqref{boundrigid} and \eqref{bound} are satisfied. Notice that this is not an artifact of the weak coupling regime, since essentially the same result is recovered near the strong coupling singularity, where the monopole becomes massless \cite{CMP}. 

This example also serves to illustrate how the asymptotic behaviour \eqref{RdivLV} is obtained within this setup. Contrary to the large volume scenario, now we have isolated states below the cut-off $m_*$. The scale above which the gauge coupling starts running is $|a|m_*$, which one identifies with the RFT cut-off $\Lambda_{\rm RFT}$. As a result, $\Lambda_g = g_{\rm rigid}^{-2}|a|m_*$, from where \eqref{RdivLV} immediately follows. 

\section{Divergences at finite volume}
Let us finally consider curvature singularities that occur at finite compactification volume in the type IIA frame. More precisely, we  consider trajectories in the vector multiplet moduli space where the (quantum) volume of the Calabi--Yau $X_3$ remains bounded, while ending at a metric singularity. One can split such trajectories into those of infinite and finite field space distance.

Infinite distance limits at finite classical overall volume have been classified in \cite{Lee:2019wij}. They all require some kind of fibration structure where the fibre, which is either a genus-one curve, a K3 or an Abelian surface, shrinks to zero size along the limit. For that reason, the leading tower predicted by the Distance Conjecture does not correspond to the D0-brane tower, but instead to some combination of branes wrapping the shrinking cycles.

Fortunately, it always happens that there exists some dual frame in which $m_*$ can be identified with the D0-brane scale, and the results obtained so far readily apply. Indeed, for K3 fibrations, quantum corrections become important and in fact prevent the quantum volume of the fibre from being taken to zero. Hence, the overall volume of the threefold must diverge along the limit, and things proceed as for Seiberg--Witten or more general emergent string singularities. In the case of an elliptic fibration, one can perform instead two T-dualities along the torus-directions (equivalently a Fourier--Mukai transform on the periods) so as to map the limit into the large volume region \cite{Corvilain:2018lgw,Lee:2019wij}, following in particular a trajectory of the form \eqref{growth}, in which $m_* \sim m_{\rm D0}$. Finally, when dealing with an Abelian surface, one may also consider performing four T-dualities along the fibre to map the trajectory to the large volume regime \cite{Lee:2019wij}. As discussed in \cite{Marchesano:2024tod}, once in that (dual) frame, the RFT sectors arise purely from fibre degenerations and those are not expected to source any curvature divergence due to their enhanced $\CN=4$ supersymmetry. 

Similarly, finite distance singularities can also be described in terms of an RFT, as we already pointed out. This time, the singularity is not generated by a divergence in $\Mpl/m_*$ but rather by a pole in the third derivative of the rigid prepotential. The latter arises from integrating out a light charged particle, whose mass becomes the RFT cut-off $\Lambda_{\rm RFT}$. Following \cite{Strominger:1995cz}, we find
\be
m_*^{-1} \mathscr{F}_{a a a} \sim \frac{1}{\pi i a}\, , \qquad g_{\rm rigid}^{-2} \sim - \frac{1}{4\pi^2} \log \left| a \right|\, ,
\label{coni}
\ee
with $a$ the hypermultiplet central charge in units of $m_*$. This results in the curvature behaving asymptotically like
\be
\mathsf{R}_{\rm rigid} \sim \frac{g_{\rm rigid}^6}{8 \pi^4 |a|^2} \implies \mathsf{R}_{\rm div} \sim \left(\frac{\Lambda_{\rm wgc}}{\Lambda_g}\right)^2 ,
\label{Rconi}
\ee
where $\Lambda_g = g^{-2}_{\rm rigid} |a|m_*$. As in the Seiberg--Witten case, we have that
\be
\gamma_{\rm hyper} \sim \frac{ \Lambda_{\rm wgc}} {|a| m_*} \gg \frac{ \Lambda_{\rm wgc}} {\Lambda_g} \, ,
\ee
since $g_{\rm rigid} \ll 1$, and so \eqref{bound} is satisfied automatically near the singularity. 	
	
\section{Conclusions}
In this work we have studied the connection between rigid field theory limits, charge-to-mass ratios $\gamma$ and field space curvature $\mathsf{R}$ divergences, in the context of type II Calabi--Yau 4d $\CN=2$ vector multiplet moduli spaces. The relation between RFT limits and some $\gamma_{\mathsf{q}}$'s that go to infinity is general, if one defines the latter as the charge-to-mass ratio of a would-be BPS particle corresponding to the quantised charges $\mathsf{q} =(p^I, q_J)$. In general, it is not always true that a given charge $\mathsf{q}$ satisfying $\gamma_{\mathsf{q}} \to \infty$ is populated by a BPS state, but this is actually the case for at least one of them in the RFT limits studied in \cite{Marchesano:2023thx,Marchesano:2024tod,CMP} and in this letter. Roughly speaking, these can be described as RFT weak coupling limits where the scalar curvature $\mathsf{R}$  diverges, or duals thereof. By looking at each of these cases individually, we have found that the curvature divergence is bounded from above by the largest charge-to-mass ratio squared, as in \eqref{bound}. This provides substantial evidence for the Curvature Criterion proposed in \cite{Marchesano:2023thx}, which states that moduli space curvature divergences can only occur when a field theory subsector decouples from gravity. 

Direct inspection of these limits in fact reveals a more accurate description of the curvature divergence, namely  \eqref{RdivLVineq}, which for geodesics should simplify to \eqref{Rdiv1}. Note that, for the RFT limits under study, this is a stronger statement than \eqref{bound}. Indeed, whenever the rigid theory is at weak coupling one expects that the lightest electrically charged particle has a mass $m_{\rm el}$ that is at or below $\Lambda_g = g_{\rm rigid}^{-2} \Lambda_{\rm RFT}$, since otherwise there would be no charged particle below this scale, going against the intuition that $\Lambda_g$ accounts for the RFT electrostatic energy. One then has that $\gamma \sim g_{\rm rigid} \Mpl/ m_{\rm el} \geq \Lambda_{\rm wcg}/ \Lambda_g$, from where \eqref{bound} follows. Moreover, as pointed out in the introduction, \eqref{Rdiv1} serves to characterise $\Lambda_{\rm RFT}$ in terms of $g_{\rm rigid}$ and $\Mpl$. Looking at the RFT limits with diverging curvature that we have analysed we find that we can separate them into two classes, depending on the nature of $\Lambda_{\rm RFT}$. In the first class, including finite-distance singularities and SW points, $\Lambda_{\rm RFT}$ is the mass of the lightest RFT-charged particle, and signals the scale where its gauge couplings start running. In the second class, including large volume limits, all RFT-charged particles are above the SDC scale $m_*$, which is identified with $\Lambda_{\rm RFT}$. Indeed, it is above $m_*$ that one sees the 4d RFT as a higer-dimensional SCFT or little string theory \cite{Marchesano:2024tod}.

Notice also that, on the one hand, \eqref{bound} is a statement that is invariant under electric-magnetic duality, and as such it is quite tempting to propose it as a general relation, valid even beyond the limits that we have studied. On the other hand, either \eqref{Rdiv1} or \eqref{RdivLVineq} have an attractive interpretation, as the decoupling of QG versus field theory UV cut-offs described in the introduction. We expect this latter condition to also have a neat duality-invariant formulation valid for more general classes of gravity-decoupling limits. Likely, such a formulation will deepen our knowledge about field space curvature divergences and rigid limits, even beyond the ones in the present framework. We hope to report further progress on this direction in the future.

\section*{Acknowledgments}
We thank Damian van de Heisteeg and Max Wiesner for helpful discussions.  This work is supported through the grants CEX2020-001007-S and PID2021-123017NB-I00, funded by MCIN/AEI/10.13039/501100011033 and by ERDF A way of making Europe. A.C. is supported by a Kadanoff as well as an Associate  KICP fellowships. L.M. is supported by the fellowship LCF/BQ/DI21/11860035 and LP by the fellowship LCF/BQ/DI22/11940039 from ``La Caixa" Foundation (ID 100010434). A. C. wishes to thank IFT-Madrid for hospitality during the different stages of this work, as well as Teresa Lobo for her continuous encouragement and support.

\bibliographystyle{apsrev4-1} 
\bibliography{papers.bib}

\begin{thebibliography}{53}%
\makeatletter
\providecommand \@ifxundefined [1]{%
 \@ifx{#1\undefined}
}%
\providecommand \@ifnum [1]{%
 \ifnum #1\expandafter \@firstoftwo
 \else \expandafter \@secondoftwo
 \fi
}%
\providecommand \@ifx [1]{%
 \ifx #1\expandafter \@firstoftwo
 \else \expandafter \@secondoftwo
 \fi
}%
\providecommand \natexlab [1]{#1}%
\providecommand \enquote  [1]{``#1''}%
\providecommand \bibnamefont  [1]{#1}%
\providecommand \bibfnamefont [1]{#1}%
\providecommand \citenamefont [1]{#1}%
\providecommand \href@noop [0]{\@secondoftwo}%
\providecommand \href [0]{\begingroup \@sanitize@url \@href}%
\providecommand \@href[1]{\@@startlink{#1}\@@href}%
\providecommand \@@href[1]{\endgroup#1\@@endlink}%
\providecommand \@sanitize@url [0]{\catcode `\\12\catcode `\$12\catcode `\&12\catcode `\#12\catcode `\^12\catcode `\_12\catcode `\%12\relax}%
\providecommand \@@startlink[1]{}%
\providecommand \@@endlink[0]{}%
\providecommand \url  [0]{\begingroup\@sanitize@url \@url }%
\providecommand \@url [1]{\endgroup\@href {#1}{\urlprefix }}%
\providecommand \urlprefix  [0]{URL }%
\providecommand \Eprint [0]{\href }%
\providecommand \doibase [0]{http://dx.doi.org/}%
\providecommand \selectlanguage [0]{\@gobble}%
\providecommand \bibinfo  [0]{\@secondoftwo}%
\providecommand \bibfield  [0]{\@secondoftwo}%
\providecommand \translation [1]{[#1]}%
\providecommand \BibitemOpen [0]{}%
\providecommand \bibitemStop [0]{}%
\providecommand \bibitemNoStop [0]{.\EOS\space}%
\providecommand \EOS [0]{\spacefactor3000\relax}%
\providecommand \BibitemShut  [1]{\csname bibitem#1\endcsname}%
\let\auto@bib@innerbib\@empty
\bibitem [{\citenamefont {Vafa}(2005)}]{Vafa:2005ui}%
  \BibitemOpen
  \bibfield  {author} {\bibinfo {author} {\bibfnamefont {C.}~\bibnamefont {Vafa}},\ }\href@noop {} {\  (\bibinfo {year} {2005})},\ \Eprint {http://arxiv.org/abs/hep-th/0509212} {arXiv:hep-th/0509212} \BibitemShut {NoStop}%
\bibitem [{\citenamefont {Arkani-Hamed}\ \emph {et~al.}(2007)\citenamefont {Arkani-Hamed}, \citenamefont {Motl}, \citenamefont {Nicolis},\ and\ \citenamefont {Vafa}}]{Arkani-Hamed:2006emk}%
  \BibitemOpen
  \bibfield  {author} {\bibinfo {author} {\bibfnamefont {N.}~\bibnamefont {Arkani-Hamed}}, \bibinfo {author} {\bibfnamefont {L.}~\bibnamefont {Motl}}, \bibinfo {author} {\bibfnamefont {A.}~\bibnamefont {Nicolis}}, \ and\ \bibinfo {author} {\bibfnamefont {C.}~\bibnamefont {Vafa}},\ }\href {\doibase 10.1088/1126-6708/2007/06/060} {\bibfield  {journal} {\bibinfo  {journal} {JHEP}\ }\textbf {\bibinfo {volume} {06}},\ \bibinfo {pages} {060} (\bibinfo {year} {2007})},\ \Eprint {http://arxiv.org/abs/hep-th/0601001} {arXiv:hep-th/0601001} \BibitemShut {NoStop}%
\bibitem [{\citenamefont {Castellano}(2024)}]{Castellano:2024bna}%
  \BibitemOpen
  \bibfield  {author} {\bibinfo {author} {\bibfnamefont {A.}~\bibnamefont {Castellano}},\ }\emph {\bibinfo {title} {{The Quantum Gravity Scale and the Swampland}}},\ \href@noop {} {Ph.D. thesis},\ \bibinfo  {school} {U. Autonoma, Madrid (main)} (\bibinfo {year} {2024}),\ \Eprint {http://arxiv.org/abs/2409.10003} {arXiv:2409.10003 [hep-th]} \BibitemShut {NoStop}%
\bibitem [{\citenamefont {Marchesano}\ \emph {et~al.}(2024{\natexlab{a}})\citenamefont {Marchesano}, \citenamefont {Melotti},\ and\ \citenamefont {Paoloni}}]{Marchesano:2023thx}%
  \BibitemOpen
  \bibfield  {author} {\bibinfo {author} {\bibfnamefont {F.}~\bibnamefont {Marchesano}}, \bibinfo {author} {\bibfnamefont {L.}~\bibnamefont {Melotti}}, \ and\ \bibinfo {author} {\bibfnamefont {L.}~\bibnamefont {Paoloni}},\ }\href {\doibase 10.1007/JHEP02(2024)103} {\bibfield  {journal} {\bibinfo  {journal} {JHEP}\ }\textbf {\bibinfo {volume} {02}},\ \bibinfo {pages} {103} (\bibinfo {year} {2024}{\natexlab{a}})},\ \Eprint {http://arxiv.org/abs/2311.07979} {arXiv:2311.07979 [hep-th]} \BibitemShut {NoStop}%
\bibitem [{\citenamefont {Marchesano}\ \emph {et~al.}(2024{\natexlab{b}})\citenamefont {Marchesano}, \citenamefont {Melotti},\ and\ \citenamefont {Wiesner}}]{Marchesano:2024tod}%
  \BibitemOpen
  \bibfield  {author} {\bibinfo {author} {\bibfnamefont {F.}~\bibnamefont {Marchesano}}, \bibinfo {author} {\bibfnamefont {L.}~\bibnamefont {Melotti}}, \ and\ \bibinfo {author} {\bibfnamefont {M.}~\bibnamefont {Wiesner}},\ }\href@noop {} {\  (\bibinfo {year} {2024}{\natexlab{b}})},\ \Eprint {http://arxiv.org/abs/2409.02991} {arXiv:2409.02991 [hep-th]} \BibitemShut {NoStop}%
\bibitem [{\citenamefont {Castellano}\ \emph {et~al.}()\citenamefont {Castellano}, \citenamefont {Marchesano},\ and\ \citenamefont {Paoloni}}]{CMP}%
  \BibitemOpen
  \bibfield  {author} {\bibinfo {author} {\bibfnamefont {A.}~\bibnamefont {Castellano}}, \bibinfo {author} {\bibfnamefont {F.}~\bibnamefont {Marchesano}}, \ and\ \bibinfo {author} {\bibfnamefont {L.}~\bibnamefont {Paoloni}},\ }\href@noop {} {\enquote {\bibinfo {title} {{To appear}},}\ }\BibitemShut {NoStop}%
\bibitem [{\citenamefont {Palti}(2017)}]{Palti:2017elp}%
  \BibitemOpen
  \bibfield  {author} {\bibinfo {author} {\bibfnamefont {E.}~\bibnamefont {Palti}},\ }\href {\doibase 10.1007/JHEP08(2017)034} {\bibfield  {journal} {\bibinfo  {journal} {JHEP}\ }\textbf {\bibinfo {volume} {08}},\ \bibinfo {pages} {034} (\bibinfo {year} {2017})},\ \Eprint {http://arxiv.org/abs/1705.04328} {arXiv:1705.04328 [hep-th]} \BibitemShut {NoStop}%
\bibitem [{\citenamefont {Heidenreich}\ \emph {et~al.}(2019)\citenamefont {Heidenreich}, \citenamefont {Reece},\ and\ \citenamefont {Rudelius}}]{Heidenreich:2019zkl}%
  \BibitemOpen
  \bibfield  {author} {\bibinfo {author} {\bibfnamefont {B.}~\bibnamefont {Heidenreich}}, \bibinfo {author} {\bibfnamefont {M.}~\bibnamefont {Reece}}, \ and\ \bibinfo {author} {\bibfnamefont {T.}~\bibnamefont {Rudelius}},\ }\href {\doibase 10.1007/JHEP10(2019)055} {\bibfield  {journal} {\bibinfo  {journal} {JHEP}\ }\textbf {\bibinfo {volume} {10}},\ \bibinfo {pages} {055} (\bibinfo {year} {2019})},\ \Eprint {http://arxiv.org/abs/1906.02206} {arXiv:1906.02206 [hep-th]} \BibitemShut {NoStop}%
\bibitem [{\citenamefont {Ferrara}\ and\ \citenamefont {Sabharwal}(1989)}]{Ferrara:1988ff}%
  \BibitemOpen
  \bibfield  {author} {\bibinfo {author} {\bibfnamefont {S.}~\bibnamefont {Ferrara}}\ and\ \bibinfo {author} {\bibfnamefont {S.}~\bibnamefont {Sabharwal}},\ }\href {\doibase 10.1088/0264-9381/6/4/002} {\bibfield  {journal} {\bibinfo  {journal} {Class. Quant. Grav.}\ }\textbf {\bibinfo {volume} {6}},\ \bibinfo {pages} {L77} (\bibinfo {year} {1989})}\BibitemShut {NoStop}%
\bibitem [{\citenamefont {Andrianopoli}\ \emph {et~al.}(1997)\citenamefont {Andrianopoli}, \citenamefont {Bertolini}, \citenamefont {Ceresole}, \citenamefont {D'Auria}, \citenamefont {Ferrara}, \citenamefont {Fre},\ and\ \citenamefont {Magri}}]{Andrianopoli:1996cm}%
  \BibitemOpen
  \bibfield  {author} {\bibinfo {author} {\bibfnamefont {L.}~\bibnamefont {Andrianopoli}}, \bibinfo {author} {\bibfnamefont {M.}~\bibnamefont {Bertolini}}, \bibinfo {author} {\bibfnamefont {A.}~\bibnamefont {Ceresole}}, \bibinfo {author} {\bibfnamefont {R.}~\bibnamefont {D'Auria}}, \bibinfo {author} {\bibfnamefont {S.}~\bibnamefont {Ferrara}}, \bibinfo {author} {\bibfnamefont {P.}~\bibnamefont {Fre}}, \ and\ \bibinfo {author} {\bibfnamefont {T.}~\bibnamefont {Magri}},\ }\href {\doibase 10.1016/S0393-0440(97)00002-8} {\bibfield  {journal} {\bibinfo  {journal} {J. Geom. Phys.}\ }\textbf {\bibinfo {volume} {23}},\ \bibinfo {pages} {111} (\bibinfo {year} {1997})},\ \Eprint {http://arxiv.org/abs/hep-th/9605032} {arXiv:hep-th/9605032} \BibitemShut {NoStop}%
\bibitem [{\citenamefont {Lauria}\ and\ \citenamefont {Van~Proeyen}(2020)}]{Lauria:2020rhc}%
  \BibitemOpen
  \bibfield  {author} {\bibinfo {author} {\bibfnamefont {E.}~\bibnamefont {Lauria}}\ and\ \bibinfo {author} {\bibfnamefont {A.}~\bibnamefont {Van~Proeyen}},\ }\href {\doibase 10.1007/978-3-030-33757-5} {\emph {\bibinfo {title} {{${\cal N}=2$ Supergravity in $D=4,5,6$ Dimensions}}}},\ Vol.\ \bibinfo {volume} {966}\ (\bibinfo {year} {2020})\ \Eprint {http://arxiv.org/abs/2004.11433} {arXiv:2004.11433 [hep-th]} \BibitemShut {NoStop}%
\bibitem [{\citenamefont {Freedman}\ and\ \citenamefont {Van~Proeyen}(2012)}]{Freedman_VanProeyen_2012}%
  \BibitemOpen
  \bibfield  {author} {\bibinfo {author} {\bibfnamefont {D.~Z.}\ \bibnamefont {Freedman}}\ and\ \bibinfo {author} {\bibfnamefont {A.}~\bibnamefont {Van~Proeyen}},\ }\href@noop {} {\emph {\bibinfo {title} {Supergravity}}}\ (\bibinfo  {publisher} {Cambridge University Press},\ \bibinfo {year} {2012})\BibitemShut {NoStop}%
\bibitem [{\citenamefont {Strominger}(1990)}]{Strominger:1990pd}%
  \BibitemOpen
  \bibfield  {author} {\bibinfo {author} {\bibfnamefont {A.}~\bibnamefont {Strominger}},\ }\href {\doibase 10.1007/BF02096559} {\bibfield  {journal} {\bibinfo  {journal} {Commun. Math. Phys.}\ }\textbf {\bibinfo {volume} {133}},\ \bibinfo {pages} {163} (\bibinfo {year} {1990})}\BibitemShut {NoStop}%
\bibitem [{\citenamefont {Craps}\ \emph {et~al.}(1997)\citenamefont {Craps}, \citenamefont {Roose}, \citenamefont {Troost},\ and\ \citenamefont {Van~Proeyen}}]{Craps:1997gp}%
  \BibitemOpen
  \bibfield  {author} {\bibinfo {author} {\bibfnamefont {B.}~\bibnamefont {Craps}}, \bibinfo {author} {\bibfnamefont {F.}~\bibnamefont {Roose}}, \bibinfo {author} {\bibfnamefont {W.}~\bibnamefont {Troost}}, \ and\ \bibinfo {author} {\bibfnamefont {A.}~\bibnamefont {Van~Proeyen}},\ }\href {\doibase 10.1016/S0550-3213(97)00408-2} {\bibfield  {journal} {\bibinfo  {journal} {Nucl. Phys. B}\ }\textbf {\bibinfo {volume} {503}},\ \bibinfo {pages} {565} (\bibinfo {year} {1997})},\ \Eprint {http://arxiv.org/abs/hep-th/9703082} {arXiv:hep-th/9703082} \BibitemShut {NoStop}%
\bibitem [{\citenamefont {Freed}(1999)}]{Freed:1997dp}%
  \BibitemOpen
  \bibfield  {author} {\bibinfo {author} {\bibfnamefont {D.~S.}\ \bibnamefont {Freed}},\ }\href {\doibase 10.1007/s002200050604} {\bibfield  {journal} {\bibinfo  {journal} {Commun. Math. Phys.}\ }\textbf {\bibinfo {volume} {203}},\ \bibinfo {pages} {31} (\bibinfo {year} {1999})},\ \Eprint {http://arxiv.org/abs/hep-th/9712042} {arXiv:hep-th/9712042} \BibitemShut {NoStop}%
\bibitem [{\citenamefont {Grimm}\ \emph {et~al.}(1978)\citenamefont {Grimm}, \citenamefont {Sohnius},\ and\ \citenamefont {Wess}}]{Grimm:1977xp}%
  \BibitemOpen
  \bibfield  {author} {\bibinfo {author} {\bibfnamefont {R.}~\bibnamefont {Grimm}}, \bibinfo {author} {\bibfnamefont {M.}~\bibnamefont {Sohnius}}, \ and\ \bibinfo {author} {\bibfnamefont {J.}~\bibnamefont {Wess}},\ }\href {\doibase 10.1016/0550-3213(78)90303-6} {\bibfield  {journal} {\bibinfo  {journal} {Nucl. Phys. B}\ }\textbf {\bibinfo {volume} {133}},\ \bibinfo {pages} {275} (\bibinfo {year} {1978})}\BibitemShut {NoStop}%
\bibitem [{\citenamefont {Breitenlohner}\ and\ \citenamefont {Sohnius}(1981)}]{Breitenlohner:1981sm}%
  \BibitemOpen
  \bibfield  {author} {\bibinfo {author} {\bibfnamefont {P.}~\bibnamefont {Breitenlohner}}\ and\ \bibinfo {author} {\bibfnamefont {M.~F.}\ \bibnamefont {Sohnius}},\ }\href {\doibase 10.1016/0550-3213(81)90470-3} {\bibfield  {journal} {\bibinfo  {journal} {Nucl. Phys. B}\ }\textbf {\bibinfo {volume} {187}},\ \bibinfo {pages} {409} (\bibinfo {year} {1981})}\BibitemShut {NoStop}%
\bibitem [{\citenamefont {Seiberg}\ and\ \citenamefont {Witten}(1994)}]{Seiberg:1994rs}%
  \BibitemOpen
  \bibfield  {author} {\bibinfo {author} {\bibfnamefont {N.}~\bibnamefont {Seiberg}}\ and\ \bibinfo {author} {\bibfnamefont {E.}~\bibnamefont {Witten}},\ }\href {\doibase 10.1016/0550-3213(94)90124-4} {\bibfield  {journal} {\bibinfo  {journal} {Nucl. Phys. B}\ }\textbf {\bibinfo {volume} {426}},\ \bibinfo {pages} {19} (\bibinfo {year} {1994})},\ \bibinfo {note} {[Erratum: Nucl.Phys.B 430, 485--486 (1994)]},\ \Eprint {http://arxiv.org/abs/hep-th/9407087} {arXiv:hep-th/9407087} \BibitemShut {NoStop}%
\bibitem [{\citenamefont {Kachru}\ \emph {et~al.}(1996)\citenamefont {Kachru}, \citenamefont {Klemm}, \citenamefont {Lerche}, \citenamefont {Mayr},\ and\ \citenamefont {Vafa}}]{Kachru:1995fv}%
  \BibitemOpen
  \bibfield  {author} {\bibinfo {author} {\bibfnamefont {S.}~\bibnamefont {Kachru}}, \bibinfo {author} {\bibfnamefont {A.}~\bibnamefont {Klemm}}, \bibinfo {author} {\bibfnamefont {W.}~\bibnamefont {Lerche}}, \bibinfo {author} {\bibfnamefont {P.}~\bibnamefont {Mayr}}, \ and\ \bibinfo {author} {\bibfnamefont {C.}~\bibnamefont {Vafa}},\ }\href {\doibase 10.1016/0550-3213(95)00574-9} {\bibfield  {journal} {\bibinfo  {journal} {Nucl. Phys. B}\ }\textbf {\bibinfo {volume} {459}},\ \bibinfo {pages} {537} (\bibinfo {year} {1996})},\ \Eprint {http://arxiv.org/abs/hep-th/9508155} {arXiv:hep-th/9508155} \BibitemShut {NoStop}%
\bibitem [{\citenamefont {Klemm}\ \emph {et~al.}(1996)\citenamefont {Klemm}, \citenamefont {Lerche}, \citenamefont {Mayr}, \citenamefont {Vafa},\ and\ \citenamefont {Warner}}]{Klemm:1996bj}%
  \BibitemOpen
  \bibfield  {author} {\bibinfo {author} {\bibfnamefont {A.}~\bibnamefont {Klemm}}, \bibinfo {author} {\bibfnamefont {W.}~\bibnamefont {Lerche}}, \bibinfo {author} {\bibfnamefont {P.}~\bibnamefont {Mayr}}, \bibinfo {author} {\bibfnamefont {C.}~\bibnamefont {Vafa}}, \ and\ \bibinfo {author} {\bibfnamefont {N.~P.}\ \bibnamefont {Warner}},\ }\href {\doibase 10.1016/0550-3213(96)00353-7} {\bibfield  {journal} {\bibinfo  {journal} {Nucl. Phys. B}\ }\textbf {\bibinfo {volume} {477}},\ \bibinfo {pages} {746} (\bibinfo {year} {1996})},\ \Eprint {http://arxiv.org/abs/hep-th/9604034} {arXiv:hep-th/9604034} \BibitemShut {NoStop}%
\bibitem [{\citenamefont {Katz}\ \emph {et~al.}(1997)\citenamefont {Katz}, \citenamefont {Klemm},\ and\ \citenamefont {Vafa}}]{Katz:1996fh}%
  \BibitemOpen
  \bibfield  {author} {\bibinfo {author} {\bibfnamefont {S.~H.}\ \bibnamefont {Katz}}, \bibinfo {author} {\bibfnamefont {A.}~\bibnamefont {Klemm}}, \ and\ \bibinfo {author} {\bibfnamefont {C.}~\bibnamefont {Vafa}},\ }\href {\doibase 10.1016/S0550-3213(97)00282-4} {\bibfield  {journal} {\bibinfo  {journal} {Nucl. Phys. B}\ }\textbf {\bibinfo {volume} {497}},\ \bibinfo {pages} {173} (\bibinfo {year} {1997})},\ \Eprint {http://arxiv.org/abs/hep-th/9609239} {arXiv:hep-th/9609239} \BibitemShut {NoStop}%
\bibitem [{\citenamefont {Billo}\ \emph {et~al.}(1998)\citenamefont {Billo}, \citenamefont {Denef}, \citenamefont {Fre}, \citenamefont {Pesando}, \citenamefont {Troost}, \citenamefont {Van~Proeyen},\ and\ \citenamefont {Zanon}}]{Billo:1998yr}%
  \BibitemOpen
  \bibfield  {author} {\bibinfo {author} {\bibfnamefont {M.}~\bibnamefont {Billo}}, \bibinfo {author} {\bibfnamefont {F.}~\bibnamefont {Denef}}, \bibinfo {author} {\bibfnamefont {P.}~\bibnamefont {Fre}}, \bibinfo {author} {\bibfnamefont {I.}~\bibnamefont {Pesando}}, \bibinfo {author} {\bibfnamefont {W.}~\bibnamefont {Troost}}, \bibinfo {author} {\bibfnamefont {A.}~\bibnamefont {Van~Proeyen}}, \ and\ \bibinfo {author} {\bibfnamefont {D.}~\bibnamefont {Zanon}},\ }\href {\doibase 10.1088/0264-9381/15/8/003} {\bibfield  {journal} {\bibinfo  {journal} {Class. Quant. Grav.}\ }\textbf {\bibinfo {volume} {15}},\ \bibinfo {pages} {2083} (\bibinfo {year} {1998})},\ \Eprint {http://arxiv.org/abs/hep-th/9803228} {arXiv:hep-th/9803228} \BibitemShut {NoStop}%
\bibitem [{\citenamefont {Gunara}\ \emph {et~al.}(2013)\citenamefont {Gunara}, \citenamefont {Louis}, \citenamefont {Smyth}, \citenamefont {Tripodi},\ and\ \citenamefont {Valandro}}]{Gunara:2013rca}%
  \BibitemOpen
  \bibfield  {author} {\bibinfo {author} {\bibfnamefont {B.~E.}\ \bibnamefont {Gunara}}, \bibinfo {author} {\bibfnamefont {J.}~\bibnamefont {Louis}}, \bibinfo {author} {\bibfnamefont {P.}~\bibnamefont {Smyth}}, \bibinfo {author} {\bibfnamefont {L.}~\bibnamefont {Tripodi}}, \ and\ \bibinfo {author} {\bibfnamefont {R.}~\bibnamefont {Valandro}},\ }\href {\doibase 10.1088/0264-9381/30/19/195014} {\bibfield  {journal} {\bibinfo  {journal} {Class. Quant. Grav.}\ }\textbf {\bibinfo {volume} {30}},\ \bibinfo {pages} {195014} (\bibinfo {year} {2013})},\ \Eprint {http://arxiv.org/abs/1305.1903} {arXiv:1305.1903 [hep-th]} \BibitemShut {NoStop}%
\bibitem [{\citenamefont {Alexandrov}\ \emph {et~al.}(2018)\citenamefont {Alexandrov}, \citenamefont {Banerjee},\ and\ \citenamefont {Longhi}}]{Alexandrov:2017mgi}%
  \BibitemOpen
  \bibfield  {author} {\bibinfo {author} {\bibfnamefont {S.}~\bibnamefont {Alexandrov}}, \bibinfo {author} {\bibfnamefont {S.}~\bibnamefont {Banerjee}}, \ and\ \bibinfo {author} {\bibfnamefont {P.}~\bibnamefont {Longhi}},\ }\href {\doibase 10.1007/JHEP01(2018)156} {\bibfield  {journal} {\bibinfo  {journal} {JHEP}\ }\textbf {\bibinfo {volume} {01}},\ \bibinfo {pages} {156} (\bibinfo {year} {2018})},\ \Eprint {http://arxiv.org/abs/1710.10665} {arXiv:1710.10665 [hep-th]} \BibitemShut {NoStop}%
\bibitem [{\citenamefont {Ceresole}\ \emph {et~al.}(1996)\citenamefont {Ceresole}, \citenamefont {D'Auria},\ and\ \citenamefont {Ferrara}}]{Ceresole:1995ca}%
  \BibitemOpen
  \bibfield  {author} {\bibinfo {author} {\bibfnamefont {A.}~\bibnamefont {Ceresole}}, \bibinfo {author} {\bibfnamefont {R.}~\bibnamefont {D'Auria}}, \ and\ \bibinfo {author} {\bibfnamefont {S.}~\bibnamefont {Ferrara}},\ }\href {\doibase 10.1016/0920-5632(96)00008-4} {\bibfield  {journal} {\bibinfo  {journal} {Nucl. Phys. B Proc. Suppl.}\ }\textbf {\bibinfo {volume} {46}},\ \bibinfo {pages} {67} (\bibinfo {year} {1996})},\ \Eprint {http://arxiv.org/abs/hep-th/9509160} {arXiv:hep-th/9509160} \BibitemShut {NoStop}%
\bibitem [{Note1()}]{Note1}%
  \BibitemOpen
  \bibinfo {note} {This relation can be satisfied exactly for certain choices of moduli fields, or to arbitrary accuracy for generic ones.}\BibitemShut {Stop}%
\bibitem [{Note2()}]{Note2}%
  \BibitemOpen
  \bibinfo {note} {For extremal BPS states, $\gamma $ corresponds to the extremality factor. Given the results of \cite {Gendler:2020dfp,Alim:2021vhs}, one does not expect RFT-charged particles to be extremal in general, so we henceforth refer to $\gamma $ as charge-to-mass ratio. It would nevertheless be interesting to compare our findings with the divergences for $\gamma $ recently found in \cite {FierroCota:2023bsp}.}\BibitemShut {Stop}%
\bibitem [{\citenamefont {Strominger}(1995)}]{Strominger:1995cz}%
  \BibitemOpen
  \bibfield  {author} {\bibinfo {author} {\bibfnamefont {A.}~\bibnamefont {Strominger}},\ }\href {\doibase 10.1016/0550-3213(95)00287-3} {\bibfield  {journal} {\bibinfo  {journal} {Nucl. Phys.}\ }\textbf {\bibinfo {volume} {B451}},\ \bibinfo {pages} {96} (\bibinfo {year} {1995})},\ \Eprint {http://arxiv.org/abs/hep-th/9504090} {arXiv:hep-th/9504090 [hep-th]} \BibitemShut {NoStop}%
\bibitem [{\citenamefont {Grimm}\ \emph {et~al.}(2018)\citenamefont {Grimm}, \citenamefont {Palti},\ and\ \citenamefont {Valenzuela}}]{Grimm:2018ohb}%
  \BibitemOpen
  \bibfield  {author} {\bibinfo {author} {\bibfnamefont {T.~W.}\ \bibnamefont {Grimm}}, \bibinfo {author} {\bibfnamefont {E.}~\bibnamefont {Palti}}, \ and\ \bibinfo {author} {\bibfnamefont {I.}~\bibnamefont {Valenzuela}},\ }\href {\doibase 10.1007/JHEP08(2018)143} {\bibfield  {journal} {\bibinfo  {journal} {JHEP}\ }\textbf {\bibinfo {volume} {08}},\ \bibinfo {pages} {143} (\bibinfo {year} {2018})},\ \Eprint {http://arxiv.org/abs/1802.08264} {arXiv:1802.08264 [hep-th]} \BibitemShut {NoStop}%
\bibitem [{\citenamefont {Grimm}\ \emph {et~al.}(2019)\citenamefont {Grimm}, \citenamefont {Li},\ and\ \citenamefont {Palti}}]{Grimm:2018cpv}%
  \BibitemOpen
  \bibfield  {author} {\bibinfo {author} {\bibfnamefont {T.~W.}\ \bibnamefont {Grimm}}, \bibinfo {author} {\bibfnamefont {C.}~\bibnamefont {Li}}, \ and\ \bibinfo {author} {\bibfnamefont {E.}~\bibnamefont {Palti}},\ }\href {\doibase 10.1007/JHEP03(2019)016} {\bibfield  {journal} {\bibinfo  {journal} {JHEP}\ }\textbf {\bibinfo {volume} {03}},\ \bibinfo {pages} {016} (\bibinfo {year} {2019})},\ \Eprint {http://arxiv.org/abs/1811.02571} {arXiv:1811.02571 [hep-th]} \BibitemShut {NoStop}%
\bibitem [{\citenamefont {Gendler}\ and\ \citenamefont {Valenzuela}(2021)}]{Gendler:2020dfp}%
  \BibitemOpen
  \bibfield  {author} {\bibinfo {author} {\bibfnamefont {N.}~\bibnamefont {Gendler}}\ and\ \bibinfo {author} {\bibfnamefont {I.}~\bibnamefont {Valenzuela}},\ }\href {\doibase 10.1007/JHEP01(2021)176} {\bibfield  {journal} {\bibinfo  {journal} {JHEP}\ }\textbf {\bibinfo {volume} {01}},\ \bibinfo {pages} {176} (\bibinfo {year} {2021})},\ \Eprint {http://arxiv.org/abs/2004.10768} {arXiv:2004.10768 [hep-th]} \BibitemShut {NoStop}%
\bibitem [{\citenamefont {Bastian}\ \emph {et~al.}(2021{\natexlab{a}})\citenamefont {Bastian}, \citenamefont {Grimm},\ and\ \citenamefont {van~de Heisteeg}}]{Bastian:2020egp}%
  \BibitemOpen
  \bibfield  {author} {\bibinfo {author} {\bibfnamefont {B.}~\bibnamefont {Bastian}}, \bibinfo {author} {\bibfnamefont {T.~W.}\ \bibnamefont {Grimm}}, \ and\ \bibinfo {author} {\bibfnamefont {D.}~\bibnamefont {van~de Heisteeg}},\ }\href {\doibase 10.1007/JHEP06(2021)162} {\bibfield  {journal} {\bibinfo  {journal} {JHEP}\ }\textbf {\bibinfo {volume} {06}},\ \bibinfo {pages} {162} (\bibinfo {year} {2021}{\natexlab{a}})},\ \Eprint {http://arxiv.org/abs/2011.08854} {arXiv:2011.08854 [hep-th]} \BibitemShut {NoStop}%
\bibitem [{\citenamefont {Ooguri}\ and\ \citenamefont {Vafa}(2007)}]{Ooguri:2006in}%
  \BibitemOpen
  \bibfield  {author} {\bibinfo {author} {\bibfnamefont {H.}~\bibnamefont {Ooguri}}\ and\ \bibinfo {author} {\bibfnamefont {C.}~\bibnamefont {Vafa}},\ }\href {\doibase 10.1016/j.nuclphysb.2006.10.033} {\bibfield  {journal} {\bibinfo  {journal} {Nucl. Phys.}\ }\textbf {\bibinfo {volume} {B766}},\ \bibinfo {pages} {21} (\bibinfo {year} {2007})},\ \Eprint {http://arxiv.org/abs/hep-th/0605264} {arXiv:hep-th/0605264 [hep-th]} \BibitemShut {NoStop}%
\bibitem [{Note3()}]{Note3}%
  \BibitemOpen
  \bibinfo {note} {As pointed out already, this may also apply for the field associated to the asymptotic trajectory, but in practice one can check that this direction does not contribute to the curvature divergence \cite {Marchesano:2023thx,Marchesano:2024tod,CMP}.}\BibitemShut {Stop}%
\bibitem [{Note4()}]{Note4}%
  \BibitemOpen
  \bibinfo {note} {Examples of these are the Seiberg-Witten points discussed in the text, or the so-called coni-LCS singularities \cite {Demirtas:2020ffz,Alvarez-Garcia:2020pxd,Bastian:2021eom}, which can also be shown to obey both \protect \eqref {Rdiv1} and \protect \eqref {bound} \cite {CMP}.}\BibitemShut {Stop}%
\bibitem [{Note5()}]{Note5}%
  \BibitemOpen
  \bibinfo {note} {The identity \protect \eqref {gammarigid} can be deduced from the no-force condition ${\protect \cal Q}_{\protect \rm rigid}^2 = G^{\mu \protect \bar {\nu }} \partial _{\mu } Z_{\protect \rm rig} \protect \bar {\partial }_{\protect \bar {\nu }} \protect \bar {Z}_{\protect \rm rig}$, which holds for BPS states in 4d ${\protect \cal N}=2$ rigid field theories.}\BibitemShut {Stop}%
\bibitem [{\citenamefont {Akhond}\ \emph {et~al.}(2022)\citenamefont {Akhond}, \citenamefont {Arias-Tamargo}, \citenamefont {Mininno}, \citenamefont {Sun}, \citenamefont {Sun}, \citenamefont {Wang},\ and\ \citenamefont {Xu}}]{Akhond:2021xio}%
  \BibitemOpen
  \bibfield  {author} {\bibinfo {author} {\bibfnamefont {M.}~\bibnamefont {Akhond}}, \bibinfo {author} {\bibfnamefont {G.}~\bibnamefont {Arias-Tamargo}}, \bibinfo {author} {\bibfnamefont {A.}~\bibnamefont {Mininno}}, \bibinfo {author} {\bibfnamefont {H.-Y.}\ \bibnamefont {Sun}}, \bibinfo {author} {\bibfnamefont {Z.}~\bibnamefont {Sun}}, \bibinfo {author} {\bibfnamefont {Y.}~\bibnamefont {Wang}}, \ and\ \bibinfo {author} {\bibfnamefont {F.}~\bibnamefont {Xu}},\ }\href {\doibase 10.21468/SciPostPhysLectNotes.64} {\bibfield  {journal} {\bibinfo  {journal} {SciPost Phys. Lect. Notes}\ }\textbf {\bibinfo {volume} {64}},\ \bibinfo {pages} {1} (\bibinfo {year} {2022})},\ \Eprint {http://arxiv.org/abs/2112.14764} {arXiv:2112.14764 [hep-th]} \BibitemShut {NoStop}%
\bibitem [{\citenamefont {Marchesano}\ and\ \citenamefont {Melotti}(2023)}]{Marchesano:2022axe}%
  \BibitemOpen
  \bibfield  {author} {\bibinfo {author} {\bibfnamefont {F.}~\bibnamefont {Marchesano}}\ and\ \bibinfo {author} {\bibfnamefont {L.}~\bibnamefont {Melotti}},\ }\href {\doibase 10.1007/JHEP02(2023)112} {\bibfield  {journal} {\bibinfo  {journal} {JHEP}\ }\textbf {\bibinfo {volume} {02}},\ \bibinfo {pages} {112} (\bibinfo {year} {2023})},\ \Eprint {http://arxiv.org/abs/2211.01409} {arXiv:2211.01409 [hep-th]} \BibitemShut {NoStop}%
\bibitem [{\citenamefont {Lanza}\ \emph {et~al.}(2021)\citenamefont {Lanza}, \citenamefont {Marchesano}, \citenamefont {Martucci},\ and\ \citenamefont {Valenzuela}}]{Lanza:2021udy}%
  \BibitemOpen
  \bibfield  {author} {\bibinfo {author} {\bibfnamefont {S.}~\bibnamefont {Lanza}}, \bibinfo {author} {\bibfnamefont {F.}~\bibnamefont {Marchesano}}, \bibinfo {author} {\bibfnamefont {L.}~\bibnamefont {Martucci}}, \ and\ \bibinfo {author} {\bibfnamefont {I.}~\bibnamefont {Valenzuela}},\ }\href {\doibase 10.1007/JHEP09(2021)197} {\bibfield  {journal} {\bibinfo  {journal} {JHEP}\ }\textbf {\bibinfo {volume} {09}},\ \bibinfo {pages} {197} (\bibinfo {year} {2021})},\ \Eprint {http://arxiv.org/abs/2104.05726} {arXiv:2104.05726 [hep-th]} \BibitemShut {NoStop}%
\bibitem [{\citenamefont {Strominger}\ \emph {et~al.}(1996)\citenamefont {Strominger}, \citenamefont {Yau},\ and\ \citenamefont {Zaslow}}]{Strominger:1996it}%
  \BibitemOpen
  \bibfield  {author} {\bibinfo {author} {\bibfnamefont {A.}~\bibnamefont {Strominger}}, \bibinfo {author} {\bibfnamefont {S.-T.}\ \bibnamefont {Yau}}, \ and\ \bibinfo {author} {\bibfnamefont {E.}~\bibnamefont {Zaslow}},\ }\href {\doibase 10.1016/0550-3213(96)00434-8} {\bibfield  {journal} {\bibinfo  {journal} {Nucl. Phys. B}\ }\textbf {\bibinfo {volume} {479}},\ \bibinfo {pages} {243} (\bibinfo {year} {1996})},\ \Eprint {http://arxiv.org/abs/hep-th/9606040} {arXiv:hep-th/9606040} \BibitemShut {NoStop}%
\bibitem [{\citenamefont {Castellano}\ \emph {et~al.}(2022)\citenamefont {Castellano}, \citenamefont {Herr\'aez},\ and\ \citenamefont {Ib\'a\~nez}}]{Castellano:2021mmx}%
  \BibitemOpen
  \bibfield  {author} {\bibinfo {author} {\bibfnamefont {A.}~\bibnamefont {Castellano}}, \bibinfo {author} {\bibfnamefont {A.}~\bibnamefont {Herr\'aez}}, \ and\ \bibinfo {author} {\bibfnamefont {L.~E.}\ \bibnamefont {Ib\'a\~nez}},\ }\href {\doibase 10.1007/JHEP08(2022)217} {\bibfield  {journal} {\bibinfo  {journal} {JHEP}\ }\textbf {\bibinfo {volume} {08}},\ \bibinfo {pages} {217} (\bibinfo {year} {2022})},\ \Eprint {http://arxiv.org/abs/2112.10796} {arXiv:2112.10796 [hep-th]} \BibitemShut {NoStop}%
\bibitem [{\citenamefont {Corvilain}\ \emph {et~al.}(2019)\citenamefont {Corvilain}, \citenamefont {Grimm},\ and\ \citenamefont {Valenzuela}}]{Corvilain:2018lgw}%
  \BibitemOpen
  \bibfield  {author} {\bibinfo {author} {\bibfnamefont {P.}~\bibnamefont {Corvilain}}, \bibinfo {author} {\bibfnamefont {T.~W.}\ \bibnamefont {Grimm}}, \ and\ \bibinfo {author} {\bibfnamefont {I.}~\bibnamefont {Valenzuela}},\ }\href {\doibase 10.1007/JHEP08(2019)075} {\bibfield  {journal} {\bibinfo  {journal} {JHEP}\ }\textbf {\bibinfo {volume} {08}},\ \bibinfo {pages} {075} (\bibinfo {year} {2019})},\ \Eprint {http://arxiv.org/abs/1812.07548} {arXiv:1812.07548 [hep-th]} \BibitemShut {NoStop}%
\bibitem [{Note6()}]{Note6}%
  \BibitemOpen
  \bibinfo {note} {When $\ker {\protect \bf K}$ is trivial, one can also have a set of diverging $\gamma $'s and an RFT-decoupling limit. However, in this case the RFT gauge couplings tend to zero quite fast, and as a result the asymptotic behaviour of the metric eigenvalues is such that no curvature divergence is generated. At most, one may have a term of the form \protect \eqref {Rdiv2} that contributes as a constant to the asymptotic curvature, rendering it in some cases positive \cite {Marchesano:2023thx}.}\BibitemShut {Stop}%
\bibitem [{\citenamefont {Lee}\ \emph {et~al.}(2022)\citenamefont {Lee}, \citenamefont {Lerche},\ and\ \citenamefont {Weigand}}]{Lee:2019wij}%
  \BibitemOpen
  \bibfield  {author} {\bibinfo {author} {\bibfnamefont {S.-J.}\ \bibnamefont {Lee}}, \bibinfo {author} {\bibfnamefont {W.}~\bibnamefont {Lerche}}, \ and\ \bibinfo {author} {\bibfnamefont {T.}~\bibnamefont {Weigand}},\ }\href {\doibase 10.1007/JHEP02(2022)190} {\bibfield  {journal} {\bibinfo  {journal} {JHEP}\ }\textbf {\bibinfo {volume} {02}},\ \bibinfo {pages} {190} (\bibinfo {year} {2022})},\ \Eprint {http://arxiv.org/abs/1910.01135} {arXiv:1910.01135 [hep-th]} \BibitemShut {NoStop}%
\bibitem [{\citenamefont {van~de Heisteeg}(2022)}]{vandeHeisteeg:2022gsp}%
  \BibitemOpen
  \bibfield  {author} {\bibinfo {author} {\bibfnamefont {D.~T.~E.}\ \bibnamefont {van~de Heisteeg}},\ }\emph {\bibinfo {title} {{Asymptotic String Compactifications: Periods, flux potentials, and the swampland}}},\ \href {\doibase 10.33540/1380} {Ph.D. thesis},\ \bibinfo  {school} {Utrecht U.} (\bibinfo {year} {2022}),\ \Eprint {http://arxiv.org/abs/2207.00303} {arXiv:2207.00303 [hep-th]} \BibitemShut {NoStop}%
\bibitem [{\citenamefont {Schmid}(1973)}]{Schmid:1973cdo}%
  \BibitemOpen
  \bibfield  {author} {\bibinfo {author} {\bibfnamefont {W.}~\bibnamefont {Schmid}},\ }\href {\doibase 10.1007/BF01389674} {\bibfield  {journal} {\bibinfo  {journal} {Invent. Math.}\ }\textbf {\bibinfo {volume} {22}},\ \bibinfo {pages} {211} (\bibinfo {year} {1973})}\BibitemShut {NoStop}%
\bibitem [{\citenamefont {Bastian}\ \emph {et~al.}(2023)\citenamefont {Bastian}, \citenamefont {Grimm},\ and\ \citenamefont {van~de Heisteeg}}]{Bastian:2021hpc}%
  \BibitemOpen
  \bibfield  {author} {\bibinfo {author} {\bibfnamefont {B.}~\bibnamefont {Bastian}}, \bibinfo {author} {\bibfnamefont {T.~W.}\ \bibnamefont {Grimm}}, \ and\ \bibinfo {author} {\bibfnamefont {D.}~\bibnamefont {van~de Heisteeg}},\ }\href {\doibase 10.1007/JHEP02(2023)149} {\bibfield  {journal} {\bibinfo  {journal} {JHEP}\ }\textbf {\bibinfo {volume} {02}},\ \bibinfo {pages} {149} (\bibinfo {year} {2023})},\ \Eprint {http://arxiv.org/abs/2108.11962} {arXiv:2108.11962 [hep-th]} \BibitemShut {NoStop}%
\bibitem [{\citenamefont {Seiberg}(1988)}]{Seiberg:1988ur}%
  \BibitemOpen
  \bibfield  {author} {\bibinfo {author} {\bibfnamefont {N.}~\bibnamefont {Seiberg}},\ }\href {\doibase 10.1016/0370-2693(88)91265-8} {\bibfield  {journal} {\bibinfo  {journal} {Phys. Lett. B}\ }\textbf {\bibinfo {volume} {206}},\ \bibinfo {pages} {75} (\bibinfo {year} {1988})}\BibitemShut {NoStop}%
\bibitem [{\citenamefont {Alim}\ \emph {et~al.}(2021)\citenamefont {Alim}, \citenamefont {Heidenreich},\ and\ \citenamefont {Rudelius}}]{Alim:2021vhs}%
  \BibitemOpen
  \bibfield  {author} {\bibinfo {author} {\bibfnamefont {M.}~\bibnamefont {Alim}}, \bibinfo {author} {\bibfnamefont {B.}~\bibnamefont {Heidenreich}}, \ and\ \bibinfo {author} {\bibfnamefont {T.}~\bibnamefont {Rudelius}},\ }\href {\doibase 10.1002/prop.202100125} {\bibfield  {journal} {\bibinfo  {journal} {Fortsch. Phys.}\ }\textbf {\bibinfo {volume} {69}},\ \bibinfo {pages} {2100125} (\bibinfo {year} {2021})},\ \Eprint {http://arxiv.org/abs/2108.08309} {arXiv:2108.08309 [hep-th]} \BibitemShut {NoStop}%
\bibitem [{\citenamefont {Fierro~Cota}\ \emph {et~al.}(2024)\citenamefont {Fierro~Cota}, \citenamefont {Mininno}, \citenamefont {Weigand},\ and\ \citenamefont {Wiesner}}]{FierroCota:2023bsp}%
  \BibitemOpen
  \bibfield  {author} {\bibinfo {author} {\bibfnamefont {C.}~\bibnamefont {Fierro~Cota}}, \bibinfo {author} {\bibfnamefont {A.}~\bibnamefont {Mininno}}, \bibinfo {author} {\bibfnamefont {T.}~\bibnamefont {Weigand}}, \ and\ \bibinfo {author} {\bibfnamefont {M.}~\bibnamefont {Wiesner}},\ }\href {\doibase 10.1007/JHEP05(2024)285} {\bibfield  {journal} {\bibinfo  {journal} {JHEP}\ }\textbf {\bibinfo {volume} {05}},\ \bibinfo {pages} {285} (\bibinfo {year} {2024})},\ \Eprint {http://arxiv.org/abs/2312.04619} {arXiv:2312.04619 [hep-th]} \BibitemShut {NoStop}%
\bibitem [{\citenamefont {Demirtas}\ \emph {et~al.}(2020)\citenamefont {Demirtas}, \citenamefont {Kim}, \citenamefont {McAllister},\ and\ \citenamefont {Moritz}}]{Demirtas:2020ffz}%
  \BibitemOpen
  \bibfield  {author} {\bibinfo {author} {\bibfnamefont {M.}~\bibnamefont {Demirtas}}, \bibinfo {author} {\bibfnamefont {M.}~\bibnamefont {Kim}}, \bibinfo {author} {\bibfnamefont {L.}~\bibnamefont {McAllister}}, \ and\ \bibinfo {author} {\bibfnamefont {J.}~\bibnamefont {Moritz}},\ }\href {\doibase 10.1002/prop.202000085} {\bibfield  {journal} {\bibinfo  {journal} {Fortsch. Phys.}\ }\textbf {\bibinfo {volume} {68}},\ \bibinfo {pages} {2000085} (\bibinfo {year} {2020})},\ \Eprint {http://arxiv.org/abs/2009.03312} {arXiv:2009.03312 [hep-th]} \BibitemShut {NoStop}%
\bibitem [{\citenamefont {\'Alvarez-Garc\'\i{}a}\ \emph {et~al.}(2020)\citenamefont {\'Alvarez-Garc\'\i{}a}, \citenamefont {Blumenhagen}, \citenamefont {Brinkmann},\ and\ \citenamefont {Schlechter}}]{Alvarez-Garcia:2020pxd}%
  \BibitemOpen
  \bibfield  {author} {\bibinfo {author} {\bibfnamefont {R.}~\bibnamefont {\'Alvarez-Garc\'\i{}a}}, \bibinfo {author} {\bibfnamefont {R.}~\bibnamefont {Blumenhagen}}, \bibinfo {author} {\bibfnamefont {M.}~\bibnamefont {Brinkmann}}, \ and\ \bibinfo {author} {\bibfnamefont {L.}~\bibnamefont {Schlechter}},\ }\href {\doibase 10.1002/prop.202000088} {\bibfield  {journal} {\bibinfo  {journal} {Fortsch. Phys.}\ }\textbf {\bibinfo {volume} {68}},\ \bibinfo {pages} {2000088} (\bibinfo {year} {2020})},\ \Eprint {http://arxiv.org/abs/2009.03325} {arXiv:2009.03325 [hep-th]} \BibitemShut {NoStop}%
\bibitem [{\citenamefont {Bastian}\ \emph {et~al.}(2021{\natexlab{b}})\citenamefont {Bastian}, \citenamefont {Grimm},\ and\ \citenamefont {van~de Heisteeg}}]{Bastian:2021eom}%
  \BibitemOpen
  \bibfield  {author} {\bibinfo {author} {\bibfnamefont {B.}~\bibnamefont {Bastian}}, \bibinfo {author} {\bibfnamefont {T.~W.}\ \bibnamefont {Grimm}}, \ and\ \bibinfo {author} {\bibfnamefont {D.}~\bibnamefont {van~de Heisteeg}},\ }\href@noop {} {\  (\bibinfo {year} {2021}{\natexlab{b}})},\ \Eprint {http://arxiv.org/abs/2105.02232} {arXiv:2105.02232 [hep-th]} \BibitemShut {NoStop}%
\end{thebibliography}%
	
\end{document}